\documentclass[a4paper,11pt,fleqn]{article}
\pdfoutput=1

\usepackage{jheppub}
\usepackage[english]{babel}
\usepackage[utf8x]{inputenc}
\usepackage{amsmath}

\newcommand{\td}{\delta}

\newcommand{\tr}{\text{tr}\,}

\makeatletter
\def\@fpheader{\relax}
\makeatother

\title{Space-time symmetries\\and the Yang-Mills gradient flow}

\author[a]{L. Del Debbio}
\author[b,c]{A. Patella}
\author[c]{A. Rago}

\affiliation[a]{Higgs Centre for Theoretical Physics, School of Physics and Astronomy, University of Edinburgh, Edinburgh EH9 3JZ, UK}
\affiliation[b]{PH-TH, CERN, CH-1211 Geneva 23, Switzerland}
\affiliation[c]{School of Computing and Mathematics \& Centre for Mathematical Science, Plymouth University, Plymouth PL4 8AA, UK}

\emailAdd{luigi.del.debbio@ed.ac.uk}
\emailAdd{agostino.patella@plymouth.ac.uk}
\emailAdd{antonio.rago@plymouth.ac.uk}

\preprint{CERN-PH-TH-2013/12}

\abstract{The recent introduction of the gradient flow has provided a
  new tool to probe the dynamics of quantum field theories. The latest
  developments have shown how to use the gradient flow for the
  exploration of symmetries, and the definition of the corresponding
  renormalized Noether currents. In this paper we introduce
  infinitesimal translations along the gradient flow for gauge
  theories, and study the corresponding Ward identities. This approach
  is readily generalized to the case of gauge theories defined on a
  lattice, where the regulator breaks translation invariance. The
  Ward identities in this case lead to a nonperturbative
  renormalization of the energy-momentum tensor. We discuss an
  application of this method to the study of dilatations and scale
  invariance on the lattice.  }


\arxivnumber{}

\begin{document}

\maketitle

\section{Introduction}

The lattice regulator provides a unique framework to investigate
non-perturbative properties of non-Abelian gauge theories. However
this formulation explicitly breaks the Poincar\'e group at finite
lattice spacing and the exact restoration of the related invariances
can be recovered only in the continuum limit.

As space-time symmetries are explicitly broken, the Ward identities
associated to translations are violated, and the construction of a
renormalized energy-momentum tensor that generates the transformations
requires special care. A nonperturbative renormalization of the
energy-momentum tensor is necessary in order to guarantee that
numerical studies of physical quantities related to the Noether
currents are not obscured by lattice artefacts. For instance the study
of scale invariance in quantum field theories is a problem that
requires the knowledge of a properly-defined energy-momentum tensor is
necessary.

The lattice energy-momentum tensor can be obtained as a linear
combination of all operators with dimension not greater than four
allowed by the lattice symmetries. The coefficients have to be tuned
in such a way that the Ward identities of the continuum are satisfied
up to cutoff effects. This condition makes sure that the defined
operator is the generator of the Poincar\'e transformations in the
continuum.  This program was articulated in great detail in
refs.~\cite{Caracciolo:1989pt, Caracciolo:1988hc}.

The approach of~\cite{Caracciolo:1989pt}
is based on the idea that one can probe the lattice energy-momentum
tensor with a certain number of local observables. However this
approach can be used only if the energy-momentum tensor is separated
from the probe observables, otherwise extra contact terms due to
mixing with higher-dimensional operators might be generated. This
problem has been occasionally seen as an intrinsic limitation of the
strategy proposed by the authors of~\cite{Caracciolo:1989pt} (see for
instance the introduction of~\cite{Suzuki:2013gza}, or the
works~\cite{Suzuki:2013gi, Suzuki:2012wx} in the context of
supersymmetry). On the contrary we argue that the limitation
originates entirely from the choice of local observables to probe
translation symmetry (or any other symmetry indeed). 

In this paper we review this program in the light of the recently
developed Yang-Mills gradient flow~\cite{Luscher:2009eq,
  Luscher:2010iy, Luscher:2011bx, Luscher:2013cpa}.  More specifically
we use the gradient flow in order to define more appropriate probes
for the translation Ward identities.  Thanks to its remarkable
renormalization properties the gradient flow offers a systematic way
to define renormalization-independent observables and finite composite
operators. The gradient flow essentially smears the elementary fields
on a typical range of order $\sqrt{8t}$ where $t$ is the flow
time. Observables constructed from the fields at some positive flow
time are non-local in the elementary fields, and they represent more
natural probes for the translation Ward identities. The main goal of
this paper is to analyse all possible divergences that can arise from
the translation Ward identities on the lattice when observables at
positive flow time are used as probes. We shall see that contact terms
are completely absent from the Ward identities, and hence they are
regular in any space-time point. In section~\ref{sec:lattice} a
strategy to renormalize the energy-momentum tensor is proposed. The
basic idea to use observables at positive flow time as probes for Ward
identities is not new, and has been already applied
in ref.~\cite{Luscher:2013cpa} to chiral symmetry.

The analysis of divergences and the regularization of Ward identities passes through a complete analysis of the space-time symmetries of the flow equation (section~\ref{sec:translations-positive}), which can be implemented as the equation of motion of a five dimensional theory~\cite{Luscher:2011bx}. Beyond the technical aspects, this analysis also generates new insight.

The Noether current associated to a symmetry is obtained by
considering some local version of the symmetry transformation. The
Ward identities describe the response to the transformation applied to
an ultra-local region of the space-time (a single point, in
distributional sense). In the case of translations the Ward identities
describe what happens if a single point of space-time gets translated
by a certain infinitesimal displacement. If a lattice regulator has
being used, this is certainly not the most natural choice. As we shall
see in section~\ref{sec:translations-positive}, the gradient flow
provides a natural way to probe symmetries at (any) intermediate length
scale, by defining quasi-local transformations, i.e. transformations
that modify the fields smoothly within a region with typical linear
size of order $\sqrt{8t}$. These quasi-local transformations do not
generate the artificial divergences arising in the ultra-local
approach, not even on the lattice.

We extend our analysis to dilatations as well
(section~\ref{sec:dilatations}). We will be able to prove an
operatorial version of the Callan-Symanzik
equation~\cite{Callan:1970yg,Symanzik:1970rt,Symanzik:1971vw}, in
which the flow time is interpreted as a (square) energy scale, and
\textit{regular} expectation values with the insertion of the trace of
the energy-momentum tensor source the violation of scale
invariance. Our analysis provides a tool to test scale-invariance at
all energy scales, which is directly related to the trace anomaly.

In a recent paper~\cite{Suzuki:2013gza}, the gradient flow is also
used to regularize the energy-momentum tensor. This very interesting
approach is orthogonal to ours: an operator is defined at each
positive flow time $t$ (and it is therefore finite in any
regularization scheme), which coincides with the energy-momentum
tensor in the $t \to 0^+$ limit. This quantity is defined in terms of
two coefficients which are calculated in a perturbative expansion. In
section~\ref{sec:small-time} we connect our general analysis with the
small flow-time expansion, and we outline a possible non-perturbative
definition of the coefficients appearing in ref.~\cite{Suzuki:2013gza}.

We also want to point out that completely different strategies have
been recently explored to renormalize the energy-momentum tensor on
the lattice~\cite{Giusti:2010bb, Giusti:2012yj}.

\section{Gradient flow --- an essential toolkit}
\label{sec:gradflow}

In this section we review the definition of the gradient flow, and
some of its salient properties that will be relevant for this
paper. Throughout this paper we use the notations of
ref.~\cite{Luscher:2011bx}. In sections~\ref{sec:translations},
\ref{sec:translations-positive}, \ref{sec:dilatations} and~\ref{sec:small-time}, we focus on
the theory on the continuum, regulated using dimensional
regularization. The dimension of space-time is taken to be
$D=4-2\epsilon$, but we will not need to use the cutoff
explicitely. In section~\ref{sec:lattice} we use an explicit lattice
discretization. In the context of lattice gauge theories, the
Yang-Mills gradient flow is referred to as Wilson flow (see
e.g.~\cite{Luscher:2010iy}), and it has been used in a number of
applications~\cite{Luscher:2010iy, Luscher:2010ik, Luscher:2010we,
  Luscher:2013cpa, Fritzsch:2013je, Fodor:2012td,
  Borsanyi:2012zr}. However we do not give a review of the Wilson flow
here, and we refer the reader to the relevant literature.

\medskip
\paragraph{Flow equations}~\\
The flow of the gauge field  $\bar{B}_{t,\mu}(x)$ is defined through
the set of equations
\begin{eqnarray}
\label{eq:flow_evolution}
&&\partial_t\bar{B}_{t,\mu} =\bar{D}_{t,\nu} \bar{G}_{t,\nu\mu} \ ,\hspace{4cm}
\left. \bar{B}_{t,\mu} \right|_{t=0} = A_\mu \ ,\\
&&\bar{G}_{t,\nu\mu} = \partial_\mu \bar{B}_{t,\nu} - \partial_\nu
\bar{B}_{t,\mu} + [\bar{B}_{t,\mu} ,
\bar{B}_{t,\nu}]
\ ,
\hspace{1cm}\bar{D}_{t,\mu}  = \partial_\mu +
[\bar{B}_{t,\mu} , \;\cdot\; ] \ ,
\end{eqnarray}
where the greek indices run only in the $D$-dimensional space, we refer to $t$
as the flow time, and $A_\mu(x) = A_\mu^A(x) T^A$ is the
fundamental gauge field of the $D$-dimensional theory.  Field
correlators involving the gauge field at flow time $t$ can be calculated
in a local field theory in $D+1$ dimensions, where the field
$B_\mu(t,x) = B^A_\mu(t,x) T^A$ is a dynamical variable, and a Lagrange multiplier
$L_\mu(t,x) = L^A_\mu(t,x) T^A$ is introduced to enforce the constraint in
eq.~(\ref{eq:flow_evolution}). The bulk action is given by: 
\begin{gather}
  \label{eq:bulk}
  S_\mathrm{bulk} = -2 \int_0^\infty dt \int d^Dx \ \tr L_\mu(t,x)
  \left\{\partial_t B_{\mu}(t,x) -D_{\nu} G_{\nu\mu}(t,x) \right\}\  .
\end{gather}
Integrating out the Lagrange multiplier $L_\mu$ yields a delta
function in the path integral
\begin{equation}
  \label{eq:deltaPI}
  \prod_{t,x} \delta\left(B_\mu(t,x)-\bar{B}_{t,\mu}(x)\right)\  ,
\end{equation}
which guarantees that the field $B_\mu(t,x)$ at flow time $t$ is the
solution of the flow equation $\bar{B}_{t,\mu}(x)$. The generators $T^A$ are antihermitean and are normalized as:
\begin{equation}
\tr T^A T^B = - \frac{1}{2} \delta_{AB} \ .
\end{equation}

A perturbative analysis of the properties of this theory has been
discussed in Ref.~\cite{Luscher:2011bx}. For our purposes, it is
interesting to emphasize that: \textit{(1)} propagators involving fields in the bulk
have an exponential suppression for large momenta, and \textit{(2)} the flow
propagator $\langle B^A_\mu(t,p) L^B_\nu(s,q)\rangle$ vanishes unless
$t>s$:
\begin{align}
  \label{eq:Bprop}
  &
  \langle B^A_\mu(t,p) B^B_\nu(s,q) \rangle =
  (2\pi)^D \delta(p+q) \delta^{AB} g_0^2 D_{t+s}(p)_{\mu\nu} + O(g_0^4) \ ,\\
  &
  D_t(p)_{\mu\nu} = \frac{1}{(p^2)^2} \left\{ \left(\delta_{\mu\nu} p^2 - p_\mu
  p_\nu\right) e^{-tp^2} + \lambda_0^{-1} p_\mu
  p_\nu e^{-\alpha_0 t p^2} \right\} \ , \\
  &
  \langle B^A_\mu(t,p) L^B_\nu(s,q)\rangle = (2\pi)^D \delta(p+q) \delta^{AB} \theta(t-s) K_{t-s}(p) + O(g_0^2)
  \\
  &
  K_t(p)_{\mu\nu} = \frac{1}{p^2} \left\{ \left(\delta_{\mu\nu} p^2 - p_\mu
  p_\nu\right) e^{-tp^2} + p_\mu
  p_\nu e^{-\alpha_0 t p^2} \right\} \ ,
\end{align}
where $\alpha_0$ and $\lambda_0$ are gauge-fixing parameters. Both
properties are useful in order to understand the structure of the
divergences in correlators involving $B$ and $L$.

\paragraph{Jacobian matrix of the trivializing map}~\\
The gradient flow is reversible, which means that the map between
field configurations at two different flow times $\bar{B}_{s} \mapsto
\bar{B}_t$ is invertible. The Jacobian matrix associated with this map
(only forward propagation is considered) is:
\begin{gather}
  J^{BA}_{\nu\mu}(t,y;s,x) = \theta(t-s) \frac{\delta
    \bar{B}^B_{t,\nu}(y)}{\delta \bar{B}^A_{s,\mu}(x)} \ .
\label{eq:jacobian}
\end{gather}
The Jacobian matrix $J$ was already introduced in ref.~\cite{Luscher:2009eq} in the context of the trivializing maps. At leading order in perturbation theory, $J$ coincides with the flow propagator:
\begin{gather}
J^{BA}_{\nu\mu}(t,y;s,x) = \delta^{AB} \theta(t-s) \int \frac{d^Dp}{(2\pi)^D} e^{ip(y-x)} K_{t-s}(p)_{\nu\mu} + O(g_0^2)\ .
\end{gather}
This Jacobian matrix has many regularity properties, of which one is
of particular interest for the discussions in later sections of this
paper. For $t>s$ the Jacobian is a regular function that decays
exponentially in $|x-y|$ as discussed e.g. in
ref.~\cite{Luscher:2011bx}. In particular at leading order in
perturbation theory it decays like $e^{-\frac{|x-y|^2}{16(t-s)}}$.

\paragraph{Local gauge-invariance in D+1 dimensions}~\\
The bulk action~\eqref{eq:bulk} is invariant under gauge
transformations that do not depend on the flow time. The bulk action
can be made invariant under local gauge transformations in $D+1$
dimensions by introducing a component $B_0$ of the gauge field along
the flow-time direction. The field-strength tensor is also extended
accordingly:
\begin{gather}
  G_{0\mu} = \partial_t B_\mu - \partial_\mu B_0 + [B_0, B_\mu]\ ,
\end{gather}
and the bulk action becomes:
\begin{gather}
  S_\mathrm{bulk} = -2 \int_0^\infty dt \int d^Dx \ \tr L_\mu(t,x)
  \left\{ G_{0\mu}(t,x) - D_\nu G_{\nu\mu}(t,x) \right\} \ .
  \label{eq:bulk-action}
\end{gather}
The original action in eq.~\eqref{eq:bulk} is recovered in the $B_0=0$
gauge. As the measure in the path integral is invariant under the
change of variables that brings to the $B_0=0$ gauge, the
actions~\eqref{eq:bulk} and~\eqref{eq:bulk-action} describe the same
quantum field theory.

\section{Translations}
\label{sec:translations}

The action of space-time translations on gauge fields can be
defined in a gauge-covariant way~\cite{Jackiw:1978ar,Berg:2000ak}:
\begin{gather}
  \label{eq:transformation_4}
  \td_\alpha A_\mu(x) \overset{\mathrm{def}}{=} \alpha_\rho(x) F_{\rho\mu}(x) \  .
\end{gather}
The associated global transformations (i.e. with a uniform
$\alpha_\rho$) reduce to infinitesimal translations up to a
field-dependent gauge transformation, and therefore are \textit{bona fide}
translations for any gauge-invariant observable. 

The four Noether currents associated with these transformations are gathered
in an energy-momentum tensor that is symmetric and gauge-invariant. If
the theory does not contain scalars this energy-momentum tensor is
uniquely determined up to the cosmological constant which we will assume
to be set equal to zero throughout this paper. For pure Yang-Mills the
energy-momentum tensor defined from the gauge-covariant
transformations above is:
\begin{gather}
  \label{eq:tmunu_4}
  T_{\mu\rho} = - \frac{2}{g_0^2} \left\{ \tr F_{\sigma\mu}
    F_{\sigma\rho} - \frac{\delta_{\mu\rho}}{4} \tr F_{\sigma\tau}
    F_{\sigma \tau} \right\} \ ,
\end{gather}
and the variation of the action under (\ref{eq:transformation_4}) is given by:
\begin{equation}
  \label{eq:deltaS}
  \delta_\alpha S = \int d^Dx\  T_{\mu\rho}(x) \partial_\mu
  \alpha_\rho(x)\ .
\end{equation}
The fields are normalized in such a way that the action takes the form:
\begin{equation}
  \label{eq:YMaction}
  S = - \frac{1}{2 g_0^2} \int d^D x \ \tr F_{\sigma\tau}
    F_{\sigma \tau}
  \ .
\end{equation}
In particular the action is invariant under the
transformation~\eqref{eq:transformation_4} when $\alpha_\rho$ is
chosen to be uniform, i.e. independent of the space-time
coordinates. It is important to stress that any explicit breaking
of the symmetry generates an extra contribution to $\delta_\alpha
S$. Such explicit breaking can originate from terms in the action, or
from the regularization used to define the theory. For instance the
lattice regularization breaks translation symmetry, leading to a
non-trivial renormalization of the energy-momentum tensor. We defer
the discussion of the broken Ward identities to
section~\ref{sec:lattice}.

The variation of a generic observable $P$ under the
transformation~\eqref{eq:transformation_4} can be written as:
\begin{gather}
  \delta_\alpha P  =
  \int d^Dx \ \alpha_\rho(x) \delta_{x,\rho} P
  \overset{\mathrm{def}}{=}
  \int d^Dx \ \alpha_\rho(x) \frac{\delta P}{\delta A^A_\mu(x)} F^A_{\rho \mu}(x)
  \ .
\end{gather}
The corresponding translation Ward identity (TWI) can be written as:
\begin{gather}
  \langle \delta_{x,\rho} P \rangle = - \langle P \partial_\mu
  T_{\mu\rho}(x) \rangle
  \ .
  \label{eq:TWI1}
\end{gather}
A more familiar form of the TWI is obtained by choosing for $P$ a
product of gauge-invariant local observables; the l.h.s. of the equation
above can be rewritten as the variation of the product of observables,
leading to: 
\begin{gather}
  \sum_j \delta(x-x_j) \frac{\partial}{\partial x_j} \langle
  \phi_1(x_1) \cdots \phi_k(x_k) \rangle = - \langle \phi_1(x_1)
  \cdots \phi_k(x_k) \partial_\mu T_{\mu\rho}(x) \rangle \ .
  \label{eq:TWI1-local}
\end{gather}
Note that the TWI~\eqref{eq:TWI1} and~\eqref{eq:TWI1-local} hold
for the regulated correlators in the bare theory, because dimensional
regularization preserves translation invariance.

A clarification is in order here. When the theory is defined using
dimensional regularization and a perturbative expansion, we must address
the issue of gauge fixing. Gauge-fixing terms and ghost terms are
added to the action, and consequently to the energy-momentum
tensor. However when gauge invariant observables are considered in the
TWI, these extra-pieces in the energy momentum tensor do
not contribute to the expectation values, so we can safely omit them.

In order to remove the cutoff in equation~\eqref{eq:TWI1-local}, the
bare parameters and fields have to be replaced with renormalized ones:
\begin{gather}
  A_\mu = Z^{1/2} Z_3^{1/2} (A_\mu)_R \ ,
  \qquad
  g_0 = \mu^{2\epsilon} g^2 Z \ ,
\end{gather}
and the observables $\phi_j$ with their renormalized counterpart
$(\phi_j)_R$:
\begin{gather}
  \label{eq:TWIforrenorm}
  \sum_j \delta(x-x_j) \frac{\partial}{\partial x_j} \langle
  \phi_1(x_1)_R \cdots \phi_k(x_k)_R \rangle = - \langle \phi_1(x_1)_R
  \cdots \phi_k(x_k)_R \partial_\mu T_{\mu\rho}(x) \rangle \ .
\end{gather}
Ward identities are a powerful tool to analyse the divergences of
Noether currents and related operators. Indeed, the finiteness (in a
distributional sense) of the l.h.s. of eq.~\eqref{eq:TWIforrenorm} in the
$\epsilon \to 0$ limit, implies the finiteness of the gauge-invariant
part of the operator $\partial_\mu T_{\mu\rho}(x)$. In gauge theories with no scalars
this is shown to be equivalent to the finiteness of the
energy-momentum tensor itself in the $\epsilon \to 0$
limit~\cite{Callan:1970ze, Adler:1976zt, Coleman:1970je,
  Collins:1976yq, Fujikawa:1980rc}. In other words the gauge-invariant
part of the energy-momentum tensor does not require renormalization in
dimensional regularization; in order to avoid the usage of an
overabundant notation we will not introduce the symbol
$(T_{\mu\nu})_R$.

For a generic non-local observable $P$, the $\epsilon \to 0$ limit of
both sides of the TWI~\eqref{eq:TWI1} is trickier because contact
terms will arise in general, and we will not pursue this direction
further. However in the next subsection we will show that, if $P$ is
chosen to be an observable that depends on the fields at positive
flow-time only, such contact terms do not arise and the corresponding
TWI is regular in the $\epsilon \to 0$ limit.

Let us conclude this introductory discussion by stressing that Ward
identities have been used routinely in the context of
renormalization. A prominent (and familiar) example of their usage is
the renormalization of quark bilinears from chiral Ward
identities~\cite{Bochicchio:1985xa}. Further progress has been made
recently in the case of the chiral Ward identities by using probe
fields at positive flow time~\cite{Luscher:2013cpa}. Following this
idea, we will investigate in the following sections the possibility of
extending the discussion of TWI at positive flow time.

\subsection{Probe observables at positive flow time}
%\label{subsec:TWI-at-positive_flowtime}

We want to specialize the TWI~\eqref{eq:TWI1} to the case of a probe
observable $P_T$ that depends only on the field $\bar{B}_{T,\mu}$ at
flow time $T>0$. The variation of $P_T$ under the
transformation~\eqref{eq:transformation_4} can be written using the
chain rule:
\begin{gather}
  \td_{x,\rho} P_T
  =
  \int d^D y \ 
  \frac{\delta P_T}{\delta \bar{B}^B_{T,\nu}(y)} J^{BA}_{\nu\mu}(T,y;0,x) F^A_{\rho \mu}(x) \ ,
\end{gather}
where the Jacobian matrix defined in eq.~\eqref{eq:jacobian} has been
used. Let us emphasise that we are considering here the variation of a
probe observable $P_T$ induced by an infinitesimal translation of the
gauge fields at 
flow time $t=0$. The expression above is purely algebraic, and it is exact for
the regulated theory, i.e. for any value of $\epsilon >0$. In order to
discuss the renormalization of the TWI, the
divergence structure of $\td_{x,\rho} P_T $ has to be understood. 
At first sight, this task seems to be difficult because the Jacobian
matrix $J$ is a non-local operator, and has a quite complicated
expansion in terms of the elementary fields of the $D$-dimensional
theory. However this problem can be completely circumvented by looking
at the extended theory in $D+1$ dimensions. Indeed let us consider the composite
operator:
\begin{gather}
  \tilde{T}_{0 \rho}(t,x) = -2 \tr L_\mu(t,x) G_{\rho \mu}(t,x) \ ,
\end{gather}
which is defined in the higher-dimensional bulk theory in terms of the
Lagrange multiplier $L_\mu$ and of the bulk field $G_{\mu\nu}$. Since
the Lagrange multiplier appears linearly in the action, any polynomial
in $L_\mu$ can be explicitly integrated out in the path integral. In
particular if the probe observable $P_T$ depends only on the field
$B_\mu(T,x)$ at flow time $T>0$ (and does not depend on the Lagrange
multiplier) it is possible to show that:
\begin{gather}
  \langle \td_{x,\rho} P_T \rangle
  =
  \langle P_T \tilde{T}_{0 \rho}(0,x) \rangle \ .
  \label{eq:variation-1}
\end{gather}
The calculation is rather technical and is reported in appendix~\ref{app:L}.
As for the case discussed above, this equation holds for any value of
$\epsilon>0$. Using eq.~\eqref{eq:variation-1} the problem of
identifying the divergences of $\langle \td_{x,\rho} P_T \rangle$ is
reduced to the standard task of identifying the divergences of the product of two operators
in the $(D+1)$-dimensional theory. We know already that no
renormalization is required for the fields in $P_T$. Also no
divergences are generated from Wick contractions of fields in $P_T$ as
the propagators of the bulk fields are exponentially suppressed at
large momenta. The same conclusion applies to Wick contractions of
fields in $\tilde{T}_{0 \rho}(0,x)$ with fields in $P_T$. Divergences can
only arise from the fact that $\tilde{T}_{0 \rho}(0,x)$ is a composite
operator of fields on the boundary. In principle $\tilde{T}_{0 \rho}(0,x)$
could mix with any other gauge-invariant operator of dimension 5 that
transforms as a vector under Lorentz transformations. However the Wick contractions
involving the Lagrange multiplier $L_\mu$ are such that the two-point
correlation functions of $\tilde{T}_{0 \rho}(0,x)$ and any local operator
composed from the gauge field at flow time zero vanish up to contact
terms. Divergent additive renormalizations to $\tilde{T}_{0 \rho}(0,x)$ by
such operators are therefore excluded. Divergences could arise from
the mixing with operators involving the Lagrange multiplier, but
$\tilde{T}_{0 \rho}(0,x)$ itself is the only one with dimension not greater
than 5 and the required symmetry properties. Therefore the operator
$\tilde{T}_{0 \rho}(0,x)$ can renormalize only multiplicatively. We
anticipate here that this argument does not rely on using dimensional
regularization, and holds also on the lattice.

In dimensional regularization, since translation invariance is
preserved, one can combine eq.~\eqref{eq:variation-1} and the
TWI~\eqref{eq:TWI1} into:
\begin{gather}
  \langle P_T \tilde{T}_{0 \rho}(0,x) \rangle
  =
  - \langle P_T \partial_\mu T_{\mu\rho}(x) \rangle
  \ .
  \label{eq:TWI2}
\end{gather}
which shows that $\tilde{T}_{0 \rho}$ stays finite in the $\epsilon
\to 0$ limit, and does not require to be renormalized. Thanks to
eq.~\eqref{eq:variation-1} the same conclusion holds for the
expectation value $\langle \td_{x,\rho} P_T \rangle$. This essentially
means that the differential operator $\td_{x,\rho}$ can at most
generate a multiplicative renormalization when applied to an
observable $P_T$ which is a function of fields at positive flow time
only, but no contact terms are generated. However the multiplicative
renormalization factor is constrained to be equal to one in
dimensional regularization thanks to translation invariance. In
section~\ref{sec:lattice} we will see how this discussion generalizes
to the case of a regularization that breaks translation invariance,
such as the lattice.

\section{Translations at positive flow time}
\label{sec:translations-positive}

The flow equations are invariant under global translations. This means
that one is free to translate the fields at any flow time $t$, the
result on any observable will be exactly the same as one would obtain
by first translating the boundary fields and then evolving them up to
flow time $t$. This argument can be taken one step further, by
generalizing the local transformation~\eqref{eq:transformation_4} as:
\begin{gather}
  \bar{\td}_{t,\alpha} P
  =
  \int d^Dx \ \alpha_\rho(x) \bar{\td}_{t,x,\rho} P
  =
  \int d^Dx \ \alpha_\rho(x) \frac{\delta P}{\delta \bar{B}^A_{t,\mu}(x)} \bar{G}^A_{t, \rho \mu}(x)
  \ .
  \label{eq:ren-transf3}
\end{gather}
This equation defines a family of transformations parametrized by the
flow time $t$. Clearly for $t=0$ the usual translation defined in the
previous section is recovered.  The differential operator
$\bar{\td}_{t,x,\rho}$ depends only on the fields $\bar{B}_{t,\mu}$ at the
space-time point $x$, but is not local in the fundamental field
$A_\mu$. As a consequence, the finite transformation generated by
$\bar{\td}_{t,x,\rho}$ modifies the fundamental field $A_\mu$ not only
at $x$, but in a neighborhood of it. This neighborhood has a typical
linear size of order $\sqrt{8t}$.

In close analogy to eq.~\eqref{eq:variation-1}, it is possible to show
that: 
\begin{equation}
  \label{eq:variation-1bis}
  \langle \bar\td_{t,x,\rho} P_T \rangle
  =
  \langle P_T \tilde{T}_{0 \rho}(t,x) \rangle \ .
\end{equation}
Note that in this case the tensor $\tilde{T}$ is evaluated at flow
time $t$, while it was computed on the boundary in eq.~\eqref{eq:variation-1}.

If $\alpha_\rho$ is uniform, the transformation generated by
$\bar{\td}_{t,x,\rho}$ reduces to the composition of a canonical
infinitesimal translation of the field $\bar{B}_{t,\mu}$ and a
field-dependent gauge transformation, which is immaterial when acting
on gauge-invariant observables. Since the flow equations are invariant
under global translations, $\bar{\td}_{t,\alpha}$
reduces to a canonical infinitesimal translation of the fields at any
flow time when acting on gauge-invariant observables:
\begin{gather}
\int d^Dy \ \bar{\td}_{t,y,\rho} \phi(x) = \partial_\rho \phi(x)
\ .
\label{eq:ren-transf4}
\end{gather}
It is interesting to consider some special instances of
eq.~\eqref{eq:ren-transf4}, e.g. by choosing an observable $\phi_T(x)$
that only depends on the field $\bar{B}_\mu$ at flow time $T$ and
space-time position $x$. If $T=t$ then a local version of
eq.~\eqref{eq:ren-transf4} holds:
\begin{gather}
\bar{\td}_{t,y,\rho} \phi_t(x) = \delta(y-x) \partial_\rho \phi_t(x)
\ .
\label{eq:ren-transf5}
\end{gather}
If $T>t$ then the delta function gets regularized and a milder result
holds. If $V$ is a sphere centered in $x$ with radius $r$ then roughly
speaking:
\begin{gather}
  \int_V d^Dy \ \bar{\td}_{t,y,\rho} \phi_T(x) = 
  \partial_\rho \phi_T(x) + O \left( e^{-\frac{r^2}{16(T-t)}} \right)
  \ .
  \label{eq:ren-transf6}
\end{gather}
We refer to the end of this section for the proof of a precise version of this equation. 

The nice feature of the differential operator $\bar{\td}_{t,\alpha}$
for $t>0$ is that it depends only on fields at positive flow time, and
therefore it does not require renormalization in any regularization
scheme. Associated with it, for each flow time $t$, there is a new
energy-momentum tensor and a new TWI. As the
transformation~\eqref{eq:ren-transf3} is non-local in the original
field $A_\mu$, this new energy-momentum tensor is not local in the
$D$-dimensional theory. However it is possible to write it in terms of
local operators in the $(D+1)$-dimensional theory by exploiting the
space-time symmetries of the $(D+1)$-dimensional theory.

\bigskip

The bulk action in eq.~\eqref{eq:bulk-action} is clearly invariant
under $(D+1)$-dimensional canonical translations (the translation in
the flow time is broken only by boundary effects). Following the
procedure described on the boundary, infinitesimal local translations
can be upgraded to the following gauge-covariant transformations
acting on the bulk fields:
\begin{gather}
\begin{cases}
\td_\alpha B_M(t,x) \overset{\mathrm{def}}{=} \alpha_R(t,x) G_{RM}(t,x)\, ,
\\
\td_\alpha L_\mu(t,x) \overset{\mathrm{def}}{=} \alpha_R(t,x) D_R L_\mu(t,x)\, ,
\end{cases}
\label{eq:transformation_5}
\end{gather}
with the constraint that $\alpha_0(0,x) = 0$. Capital indices run from
$0$ to $D$, and the index $0$ denotes the flow time. We will always
consider here observables that do not depend on the Lagrange
multiplier $L_\mu$. The variation of one of these observables $P$ is:
\begin{flalign}
\td_\alpha P
& \overset{\phantom{\mathrm{def}}}{=} 
\int_0^\infty dt \int d^Dx \ \alpha_R(t,x) \td_{t,x,R} P
\overset{\mathrm{def}}{=}
\nonumber \\
& \overset{\mathrm{def}}{=}
\int_0^\infty dt \int d^Dx \ \alpha_R(t,x) \frac{\delta P}{\delta B^A_M(t,x)} G^A_{RM}(t,x)
\ .
\end{flalign}

The variation of the bulk action under the transformation~\eqref{eq:transformation_5} defines a $(D+1)$-dimensional energy-momentum tensor:
\begin{gather}
\td_\alpha S_\mathrm{bulk} = \int_0^\infty dt \int d^Dx \ \tilde{T}_{MR}(t,x) \partial_M \alpha_R(t,x)
\ ,
\\
\tilde{T}_{0 R} = -2 \tr L_\mu G_{R \mu}
\ ,
\label{eq:T_0R}
\\
\tilde{T}_{\nu R} =
2 \tr L_\mu D_\nu G_{R\mu} 
-2 \tr D_\nu L_\mu G_{R\mu}
-2 \tr D_\mu L_\mu G_{\nu R}
+ 2 \delta_{R0} \tr L_\nu D_\mu G_{\mu 0}
\ ,
\label{eq:T_nR}
\end{gather}
up to terms that are proportional to the constraint and therefore
vanish in expectation values. Notice that the operator $\tilde{T}_{0
  R}$ for $R \neq 0$ is the same that appears in eqs.~\eqref{eq:variation-1} and
\eqref{eq:TWI2}.
As the number of differential operators is proliferating, we find convenient to review at this point the meaning of all of them. The differential operator $\bar{\delta}_{t,x,\rho}$ acts on fields that satify already the flow equation. The fields are deformed at flow time $t$, and the flow equation propagates this deformation to all other flow times. To make sense of this picture, we use the fact that the flow is invertible at least at finite cutoff. In particular the operator $\delta_{x,\rho} = \bar{\delta}_{0,x,\rho}$ deforms the fields on the boundary, i.e. the initial condition for the flow equation, and therefore the deformation is propagated to any positive flow time. The operator $\delta_{t,x,\rho}$ that we have just defined is completely different, as it acts on the $(D+1)$-dimensional fields \textit{before} the flow equation is imposed. It deforms the fields locally in the $(D+1)$-dimensional space and such deformation is not propagated in flow time. Of course if one starts with a field configuration that satisfies the flow equation, its deformation will in general not satisfy the same equation. The variation in the equation is reabsorbed by the deformation of the Lagrange multiplier.

For any
$t>0$, the Ward identities associated with the transformation \eqref{eq:transformation_5} are:
\begin{gather}
  \label{eq:5dimTWI}
  \langle \td_{t,x,\rho} P \rangle = - \langle P \partial_M \tilde{T}_{M R}(t,x) \rangle \ .
\end{gather}
For a probe observable $P_T$ that depends only on the field $B_\mu$ at
flow time $T>t$, the l.h.s. of the previous equation vanishes. In
this particular case, eq.~\eqref{eq:5dimTWI}  can be written as:
\begin{gather}
  \langle P_T \partial_t \tilde{T}_{0 R}(t,x) \rangle = - \langle
  P_T \partial_\mu \tilde{T}_{\mu R}(t,x) \rangle \ .
  \label{eq:main-equation}
\end{gather}
We will not need to consider the case $R=0$ in this
section, and we will therefore develop the arguments below for the
case where $R=\rho$ spans the usual space-time directions. 
We will see now how
eq.~\eqref{eq:main-equation} leads to the Ward identities for the family of
transformations defined in eq.~\eqref{eq:ren-transf3}, and how one can
use this equation to prove eq.~\eqref{eq:ren-transf6}. Note that all
fields that are computed at flow
time $t>0$ have finite correlators, and do not
require renormalization as the regularization is removed. 

\bigskip

We would like to integrate eq.~\eqref{eq:main-equation} in flow time
in an interval $(0,t)$. However this equation is valid only at
positive flow time. The problem is that for $t=0$ the Ward
identity~\eqref{eq:5dimTWI} gets an extra contribution from the fact
that the boundary fields are transformed along with the bulk
ones. Moreover eq.~\eqref{eq:main-equation} is valid for bare fields
at finite cutoff. At positive $t$, since only fields in the bulk are
involved, this equation does not have any divergences and its
$\epsilon \to 0$ limit can be safely taken. Therefore, after the
cutoff is removed, we integrate eq.~\eqref{eq:main-equation} in an
interval $(t_0,t)$ first with $0<t_0<t<T$:
\begin{gather}
  \langle P_T \tilde{T}_{0 \rho}(t,x) \rangle = \langle P_T \tilde{T}_{0
    \rho}(t_0,x) \rangle - \langle P_T \partial_\mu \int_{t_0}^t ds \
  \tilde{T}_{\mu \rho}(s,x) \rangle \ ,
  \label{eq:integrated-main-equation-regular}
\end{gather}
and then we take the $t_0 \to 0^+$ limit. We have
already proven eq.~\eqref{eq:variation-1bis}:
\begin{equation}
  \label{eq:repetita}
  \langle P_T \tilde{T}_{0\rho}(t,x)\rangle = 
  \langle\bar{\td}_{t,x,\rho} P_T\rangle \ ,
\end{equation}
and eq.~\eqref{eq:TWI2}:
\begin{gather}
  \label{eq:juvant}
  \lim_{t_0 \to 0^+} \langle P_T \tilde{T}_{0 \rho}(t_0,x) \rangle = -
  \langle P_T \partial_\mu T_{\mu\rho}(x) \rangle \ .
\end{gather}
By using these results, eq.~\eqref{eq:integrated-main-equation-regular} can be repackaged into the TWI associated with the differential operator $\bar{\td}_{t,x,\rho}$, which defines the corresponding energy momentum tensor $\bar{T}_{\mu \rho}$:
\begin{gather}
  \langle \bar{\td}_{t,x,\rho} P_T \rangle = - \langle
  P_T \partial_\mu \bar{T}_{\mu \rho}(t,x) \rangle \ ,
  \label{eq:TWI3}
  \\
  \bar{T}_{\mu \rho}(t,x) = T_{\mu \rho}(x) + \int_{0}^t ds \ \tilde{T}_{\mu \rho}(s,x) 
  \ .
  \label{eq:TWI3-barT}
\end{gather}
Clearly this TWI reduces to eq.~\eqref{eq:TWI1} for $t=0$; however
these manipulations are meaningful only if the  integral
appearing in the energy-momentum tensor \eqref{eq:TWI3-barT} is
finite.

The possible divergences of $\tilde{T}_{\mu \rho}(s,x)$ at $s \to 0^+$ are classified in terms of all operators of dimension up to 6 that can mix with $\tilde{T}_{\mu \rho}(s,x)$. Such
operators must contain at least a Lagrange multiplier, i.e. an operator of
dimension 3. Therefore, by taking into account the Lorentz structure, $\tilde{T}_{\mu \rho}(s,x)$ can mix with operators of dimension 6 and 4. However it is easy to see that gauge-invariance excludes operators of dimension 4. This means that $\tilde{T}_{\mu \rho}(s,x)$ has at most a logarithmic divergence for $s \to 0^+$, which is integrable. This concludes our discussion,
as the singularity in the energy-momentum tensor \eqref{eq:TWI3-barT} is integrable.

\bigskip

In order to understand the action of the operator $\bar{\td}_{t,x,\rho}$ on
fields defined at $T>t$, let us now integrate
eq.~\eqref{eq:main-equation} in flow time in the interval
$(t,T)$. Using eq.~\eqref{eq:variation-1bis} again:
\begin{gather}
\langle \bar{\td}_{t,x,\rho} P_T \rangle = \langle \bar{\td}_{T,x,\rho} P_T \rangle + \langle P_T \partial_\mu \int_t^T ds \ \tilde{T}_{\mu \rho}(s,x) \rangle
\ .
\end{gather}
Let us consider a local observable $\phi(T,x)$ at positive flow time
$T$, and let $X_T$ be a product of other local observables at the same
flow time $T$ but different space-time positions. We choose $P_T = X_T
\phi(T,x)$ in the previous equation and integrate it on a space-time
sphere $V$ with radius $r$ and centered in $x$. Assuming that all the
local observables in $X_T$ lie outside of the sphere $V$, and by using
eq.~\eqref{eq:ren-transf5}, one gets:
\begin{flalign}
& \langle  X_T \int_V d^Dy \ \bar{\td}_{t,y,\rho} \phi(T,x) \rangle = \nonumber \\
& \qquad =
\langle  X_T \partial_\rho \phi(T,x) \rangle
+ \langle X_T \phi(T,x) \int_t^T ds \int_{\partial V} dS_\mu \ \tilde{T}_{\mu \rho}(s,x) \rangle
\ .
\end{flalign}
The operator $\tilde{T}_{\mu \rho}$ contains only terms that are
linear in the Lagrange multiplier $L_\mu$. Since the propagator $LB$
is exponentially suppressed with the space-time separation of the two
fields, the contribution of the last term of the previous equation is
exponentially suppressed if all the fields are far enough from the
boundary $\partial V$ of the sphere. If $\bar{r}$ is distance from
$\partial V$ of the closest operator (clearly $\bar{r} \le r$), then:
\begin{gather}
\langle  X_T \int_V d^Dy \ \bar{\td}_{t,y,\rho} \phi(T,x) \rangle =
\langle  X_T \partial_\rho \phi(T,x) \rangle
 + O \left( e^{-\frac{\bar{r}^2}{16(T-t)}} \right)
\ ,
\label{eq:ren-transf6-bis}
\end{gather}
which is the precise form of eq.~\eqref{eq:ren-transf6}.

\section{Dilatations}
\label{sec:dilatations}

In order to discuss dilatations, we are going to extend the definition of
the differential operator $\bar{\td}_{t,x,\rho}$ in
eq.~\eqref{eq:ren-transf3} to include the flow-time direction:
\begin{gather}
  \bar{\td}_{t,x,R} P
  \overset{\mathrm{def}}{=}
  \frac{\delta P}{\delta \bar{B}^A_{t,\mu}(x)} \bar{G}^A_{t, R \mu}(x)
  \ ,
  \label{eq:ren-transf7}
\end{gather}
where $R$ runs over all $D+1$ dimensions. 
This differential operator can be related to the $\tilde{T}_{0 R}$ operator at generic flow time:
\begin{gather}
  \langle \bar{\td}_{t,x,R} P_T \rangle
  =
  \langle P_T \tilde{T}_{0 R}(t,x) \rangle \ ,
  \label{eq:variation-2}
\end{gather}
by integrating explicitly the Lagrange multiplier, as shown in appendix~\ref{app:L}.

Local dilatations are a special case of local translations. On the
boundary a local dilatation is generated by the
transformation~\eqref{eq:transformation_4} with $\alpha_\rho(x) =
x_\rho \beta(x)$. A global dilatation corresponds to a uniform
$\beta$. The flow equation is also invariant under dilatations
provided that the flow time is rescaled too by its classical
dimension. Local dilatations in the bulk are generated by the
transformation~\eqref{eq:transformation_5} with $\alpha_\rho(t,x) =
x_\rho \beta(t,x)$ and $\alpha_0(t,x) = 2t \beta(t,x)$.

In practice we consider the equation:
\begin{gather}
\langle P_T \partial_t \left[ 2 t \, \tilde{T}_{00}(t,x) + x_\rho \, \tilde{T}_{0\rho}(t,x) \right] \rangle = \nonumber \\
= \langle P_T \left[ 2 \tilde{T}_{00}(t,x) + \tilde{T}_{\mu\mu}(t,x) \right] \rangle
- \langle P_T \partial_\mu \left[ 2t \, \tilde{T}_{\mu 0}(t,x) + x_\rho \, \tilde{T}_{\mu\rho}(t,x) \right] \rangle \ ,
\label{eq:dilatations-1}
\end{gather}
which follows trivially from eq.~\eqref{eq:main-equation} and stays
finite in the $\epsilon \to 0$ limit. It is interesting to notice that
the operator $2 \tilde{T}_{00} + \tilde{T}_{\mu\mu}$ (which is almost
the trace of the bulk energy-momentum tensor, except that different
components are weighted with the dimension of the corresponding
coordinate) might break dilatation invariance in the bulk. However
some trivial algebra shows that this generalized trace is a
divergence:
\begin{gather}
2 \tilde{T}_{00}(t,x) + \tilde{T}_{\mu\mu}(t,x) = \partial_\mu \tilde{T}_{0\mu}(t,x) \ ,
\label{eq:dilatations-2}
\end{gather}
up to terms that are proportional to the constraint generating the flow equation, which we can omit as they vanish in expectation values. This result is not surprising as the flow equation is invariant under dilatations. We plug this result back into eq.~\eqref{eq:dilatations-1}, and integrate it in the flow-time interval $(0,t)$ with $t<T$, following closely what we have done already for the TWI:
\begin{gather}
\langle P_T \left[ 2 t \, \tilde{T}_{00}(t,x) + x_\rho \, \tilde{T}_{0\rho}(t,x) \right] \rangle
-
\lim_{t_0 \to 0^+} \langle P_T \left[ 2 t_0 \, \tilde{T}_{00}(t_0,x) + x_\rho \, \tilde{T}_{0\rho}(t_0,x) \right] \rangle
= \nonumber \\
=
- \langle P_T \partial_\mu \int_{0}^t ds \ \left[ 2s \, \tilde{T}_{\mu 0}(s,x) + x_\rho \, \tilde{T}_{\mu\rho}(s,x) - \tilde{T}_{0\mu}(s,x)\right] \rangle \ ,
\label{eq:dilatations-3}
\end{gather}
Now we can use eqs.~\eqref{eq:variation-2} and \eqref{eq:TWI2}, together with the observation that:
\begin{gather}
\lim_{t_0 \to 0^+} \langle P_T t_0 \, \tilde{T}_{00}(t_0,x) \rangle = 0 \ ,
\end{gather}
as $\tilde{T}_{00}(t_0,x)$ diverges at most logarithmically, and we can repackage eq.~\eqref{eq:dilatations-3} into the dilatation Ward identity (DWI):
\begin{gather}
  \langle \left\{ 2 t \bar{\td}_{t,x,0} + x_\rho \bar{\td}_{t,x,\rho}
  \right\} P_T \rangle = - \langle P_T \partial_\mu \bar{D}_{\mu}(t,x)
  \rangle + \langle P_T T_{\mu\mu}(x) \rangle \ ,
  \label{eq:DWI}
  \\
  \bar{D}_{\mu}(t,x) = x_\rho T_{\mu\rho}(x) + \int_{0}^t ds \
  \tilde{D}_{\mu}(s,x) \ ,
  \\
  \tilde{D}_\mu = 2t\, \tilde{T}_{\mu 0} + x_\rho \, \tilde{T}_{\mu \rho} -
  \tilde{T}_{0 \mu} \ ,
\end{gather}
where $2 t \bar{\td}_{t,x,0} + x_\rho \bar{\td}_{t,x,\rho}$ is the
differential operator that generates dilatations at flow time $t$. Usual power-counting arguments show that the integral in the dilatation current $\bar{D}_{\mu}$ is finite.
As
usual $P_T$ is an observable that depends on the field $B_\mu$ at flow
time $T$ only, and $T>t$. Of course dilatations are not symmetries of
pure Yang-Mills. The trace of the energy-momentum tensor that appears
in the r.h.s. of the DWI~\eqref{eq:DWI} is the source of the anomaly.

If $\phi_T(x)$ is an observable that depends only on the field
$\bar{B}_T$ at flow time $T$ and space-time point $x$, then the
global dilatation is simply:
\begin{gather}
  \int d^Dy \ \left\{ 2 t \bar{\td}_{t,y,0} + y_\rho
    \bar{\td}_{t,y,\rho} \right\} \phi_T(x) = \left\{ 2 T \frac{d}{dT}
    + x_\rho \partial_\rho + d_\phi\right\} \phi_T(x) \ ,
\end{gather}
where $d_\phi$ is the dimension of the operator $\phi_T$. The DWI for
$\phi_T$ reduces to a very simple form:
\begin{gather}
  \left( 2 T \frac{d}{dT} + d_\phi\right) \langle \phi_T \rangle =
  \langle \phi_T \int d^Dx \ T_{\mu\mu}(x) \rangle_c \ .
\label{eq:CS}
\end{gather}
This equation is the operatorial form of the Callan-Symanzik
equation~\cite{Callan:1970yg,Symanzik:1970rt,Symanzik:1971vw}, in
which $(8T)^{-1/2}$ is the energy scale, and contact terms are absent
(which is the same as saying that the operator $\phi_T$ does not
renormalize). Equation~\eqref{eq:CS} is extremely interesting as it
allows the trace of the energy-momentum tensor to b probed just by
looking at the evolution under gradient flow of observables.

\section{Space-time symmetries on the lattice}
\label{sec:lattice}

If the lattice regulator is used, then the explicit breaking of
translation symmetry generates an extra term in the
TWI~\eqref{eq:TWI1}, which implies that the energy-momentum tensor
will require renormalization. Even after subtracting the divergences,
the TWI~\eqref{eq:TWI1} is valid in this case only up to terms that
vanish in infinite-cutoff limit. We will review how this happens,
following the presentation in ref.~\cite{Caracciolo:1989pt}.

At finite lattice spacing $a$, a regularized version of the
transformation~\eqref{eq:transformation_4}, can be defined by choosing
for example a particular discretization of the field strength
$F_{\mu\nu}$ (for definiteness one can adopt the clover plaquette
definition) and by replacing the fundamental field $A_\mu(x)$ with the
link variable $U_\mu(x)$. The discretized transformation will be
denoted by $\hat{\td}$:
\begin{gather}
  \hat{\td}_\alpha U_\mu(x) = \alpha_\rho(x) \hat{F}_{\rho\mu}(x) U_\mu(x)
  \ ,
  \\
  \hat{\td}_{x,\rho} P
  =
  \frac{1}{a^3} \hat{F}^A_{\rho \mu}(x) \partial^A_{U_\mu(x)} P
  \ ,
  \label{eq:transformation-lat}
\end{gather}
where $\partial^A_{U_\mu(x)}$ is the left Lie derivative on the gauge
group with respect to $U_\mu(x)$. This transformation leaves the
measure of the path integral unchanged, however it is not a symmetry
as the lattice action $\hat{S}$ is not invariant when the parameter
$\alpha_\rho$ is chosen to be uniform:
\begin{gather}
  \hat{\td}_\alpha \hat{S} = - \sum_x a^4 \alpha_\rho(x) \left\{
    R_\rho(x) + \hat{\partial}_\mu \hat{T}^{(1)}_{\mu\rho}(x) \right\}
  \ ,
  \label{eq:deltaS-lat}
\end{gather}
where $\hat{T}^{(1)}_{\mu\rho}$ is your favourite naively-discretized
energy-momentum tensor. The $R_\rho$ operator, which depends on the
choice of discretization for $\hat{T}^{(1)}_{\mu\rho}$, is the residual term in
the Ward identity, and comes from the explicit breaking of the
symmetry. It is a higher-dimensional operator, and vanishes in the
formal $a \to 0$ limit (i.e. on fixed field configurations that have a
smooth continuum limit). However formally subleading corrections
cannot be neglected in field correlators as subleading coefficients
can combine with divergent expectation values giving rise to finite
contributions. By standard dimensional analysis arguments one can
isolate the possible divergences in $R_\rho$:
\begin{gather}
  R_\rho = \frac{1}{Z_\td} \bar{R}_\rho + \left( \frac{c_1}{Z_\td} - 1
  \right) \hat{\partial}_\mu \hat{T}^{(1)}_{\mu\rho} +
  \frac{c_2}{Z_\td} \hat{\partial}_\mu \hat{T}^{(2)}_{\mu\rho} +
  \frac{c_3}{Z_\td} \hat{\partial}_\mu \hat{T}^{(3)}_{\mu\rho} \ ,
\end{gather}
where $\bar{R}_\rho$ is a finite operator, and the $\hat{T}^{(2,3)}_{\mu\rho}$ operators are:
\begin{gather}
  \hat{T}^{(2)}_{\mu\rho} = \delta_{\mu\rho} \sum_{\sigma\tau} \tr \hat{F}_{\sigma\tau} \hat{F}_{\sigma \tau}
  \ ,
  \\
  \hat{T}^{(3)}_{\mu\rho} = \delta_{\mu\rho} \tr \hat{F}_{\mu\rho} \hat{F}_{\mu\rho}
  \ .
\end{gather}
If the renormalized energy-momentum tensor on the lattice is defined as:
\begin{gather}
  (\hat{T}_{\mu\rho})_R = \sum_i c_i \left\{ \hat{T}^{(i)}_{\mu\rho} -
    \langle \hat{T}^{(i)}_{\mu\rho} \rangle \right\} \ ,
\end{gather}
the Ward identity associated with the transformation~\eqref{eq:transformation-lat} becomes:
\begin{gather}
\langle Z_\td \hat{\td}_{x,\rho} P + P \bar{R}_\rho(x) \rangle =
- \langle P \hat{\partial}_\mu \hat{T}_{\mu\rho}(x)_R \rangle \ .
\label{eq:lat-R1}
\end{gather}
As for the case of dimensional regularization, we will discuss the
continuum limit of this equation for two possible choices of the
observable $P$: a product of local observables, or a generic
observable that depends on the fields at positive flow time only.

\bigskip

Let us choose for $P$ a product of properly renormalized local observables at separate points. The assumption that translation symmetry has to be recovered in the continuum limit implies that the coefficients $c_i$ and $Z_\delta$ can be tuned in such a way that: \textit{(1)} the energy-momentum tensor is finite in the continuum limit, i.e. the following limit:
\begin{gather}
  \lim_{a \to 0} \langle \hat{\phi}_1(x_1)_R \cdots \hat{\phi}_k(x_k)_R \hat{T}_{\mu\rho}(x)_R \rangle
  =
  \langle \phi_1(x_1)_R \cdots \phi_k(x_k)_R T_{\mu\rho}(x) \rangle
  \label{eq:lat-TWI-rhs}
\end{gather}
is finite up to contact terms; \textit{(2)} the l.h.s. of
eq.~\eqref{eq:lat-R1} is finite in the continuum limit in
distributional sense, and is equal to:
\begin{flalign}
& \lim_{a \to 0} \langle \left\{ Z_\td \hat{\td}_{x,\rho} +
\bar{R}_\rho(x) \right\} \hat{\phi}_1(x_1)_R \cdots
\hat{\phi}_k(x_k)_R \rangle = \nonumber \\
& \qquad = \sum_j \delta(x-x_j)
\frac{\partial}{\partial x_j} \langle \phi_1(x_1)_R \cdots
\phi_k(x_k)_R \rangle \ .
\label{eq:lat-TWI-lhs}
\end{flalign}
In particular this implies that the term $\langle \bar{R}_\rho(x)
\hat{\phi}_1(x_1)_R \cdots \hat{\phi}_k(x_k)_R \rangle$ is zero in the
continuum limit up to contact terms. These contact terms cancel
analogous contact terms arising in $Z_\td \langle \hat{\td}_{x,\rho}
\hat{\phi}_1(x_1)_R \cdots \hat{\phi}_k(x_k)_R \rangle$. However some
divergences in the cutoff have to survive in
eq.~\eqref{eq:lat-TWI-lhs} in order to reproduce the delta
function. These divergences have a purely algebraic origin in the
continuum limit, and show that local operators do not necessarily
represent the most natural choice to probe the translation Ward
identity.

\bigskip

Let us consider now a probe observable $\hat{P}_T$ which is function
of the fields at positive flow time $T$ only. As in dimensional
regularization, also the lattice differential operator
$\hat{\td}_{x,\rho}$ can be represented by a local operator in the
$(D+1)$-dimensional theory. As discussed in the
appendix~\ref{app:L-lat}, the following equality holds at any lattice
spacing:
\begin{gather}
\langle \hat{\td}_{x,\rho} \hat{P}_T \rangle
=
-2 \langle \hat{P}_T \tr L_\mu(0,x) \hat{F}_{\rho\mu}(x) \rangle \ ,
\label{eq:variation-1-lat}
\end{gather}
which is the discretized version of eq.~\eqref{eq:variation-1}. As discussed in section~\ref{sec:translations-positive} the operator $\tr L_\mu(0,x) \hat{F}_{\rho\mu}(x)$ renormalizes multiplicatively. One can therefore introduce the renormalized operator:
\begin{gather}
\tilde{T}_{0\rho}(0,x)_R
=
-2 Z_\delta \tr L_\mu(0,x) \hat{F}_{\rho\mu}(x) \ ,
\label{eq:reonrmalized_Ttilde}
\end{gather}
such that the limits that appear in the following chain of equations are finite:
\begin{gather}
\lim_{a \to 0} \langle \hat{P}_T \tilde{T}_{0\rho}(0,x)_R \rangle
=
\lim_{a \to 0} Z_\delta \langle \hat{\td}_{x,\rho} \hat{P}_T \rangle
=
\langle \td_{x,\rho} P_T \rangle \ .
\label{eq:variation-2-lat}
\end{gather}
As the operator $\tilde{T}_{0\rho}(0,x)$ is renormalization group
invariant in the continuum, the renormalization of the corresponding
lattice-discretized operator is finite, i.e. $Z_\delta$ is depends on
the lattice spacing only through the bare coupling. This finite
normalization must be fixed by requiring that the continuum
differential operator $\td_{x,\rho}$ defined through
eq.~\eqref{eq:variation-2-lat} generates translations, or in other
words satisfies eq.~\eqref{eq:ren-transf6} for $t=0$. It is important
to notice also that no contact term is generated in
eq.~\eqref{eq:variation-2-lat}, therefore its continuum limit is
regular as a function of the space-time position. Roughly speaking, through
eq.~\eqref{eq:variation-2-lat} the use of observables at positive flow
time allows the renormalization of the differential operator
$\hat{\td}_{x,\rho}$ without using the assumption that translation
invariance must be recovered in the continuum limit. Under such
supplementary assumption one concludes that the coefficients $c_i$ can
be tuned in such a way that: \textit{(1)} the energy-momentum tensor
is finite in the continuum limit, i.e. the following limit:
\begin{gather}
\lim_{a \to 0} \langle \hat{P}_T \hat{T}_{\mu\rho}(x)_R \rangle
=
\langle P_T T_{\mu\rho}(x) \rangle
\end{gather}
is finite and regular at the space-time point $x$ (as no contact terms are generated); \textit{(2)} the contribution of the operator $\bar{R}$ vanishes in the continuum limit at any space-time point $x$:
\begin{gather}
\lim_{a \to 0} \langle \hat{P}_T \bar{R}_\rho(x) \rangle
= 0 \ .
\end{gather}

Putting all together, up to subleading corrections in the lattice spacing:
\begin{gather}
\langle \td_{x,\rho} P_T \rangle = Z_\td \langle \hat{\td}_{x,\rho} \hat{P}_T \rangle = - \langle \hat{P}_T \hat{\partial}_\mu \hat{T}_{\mu\rho}(x)_R \rangle \ .
\label{eq:lat-TWI-t}
\end{gather}
Probe observables defined in terms of the fields at some positive flow time generate neither delta functions in the Ward identity nor contact terms, and seem to represent a more natural choice to probe translation symmetry (or any other symmetry).

\subsection{Strategies to renormalize the energy-momentum tensor}

Let us consider a local observable $\phi_{t,\rho}(x)$ in the fields at positive flow time $t$. For reasons that will be clear soon, we choose it to transform like a vector with respect to the hypercubic symmetry. Up to subleading corrections in the lattice spacing, eq.~\eqref{eq:lat-TWI-t} implies:
\begin{gather}
Z_\td \langle \hat{\td}_{x,\rho} \phi_{t, \rho}(0) \rangle
= -
\sum_i c_i \langle \phi_{t,\rho}(0) \hat{\partial}_\mu \hat{T}^{(i)}_{\mu\rho}(x)  \rangle
\ .
\label{eq:lat-proposal-1}
\end{gather}
In the continuum limit this equation is valid for any space-time position $x$ (even $x=0$), any flow time $t$, and any probe observable. The ratios $c_i/Z_\td$ are therefore highly constrained by this equation. We have chosen a vector probe so that the expectation values in eq.~\eqref{eq:lat-proposal-1} do not vanish at $x=0$.

\bigskip

We need to fix now the multiplicative renormalization $Z_\td$. This can be done in several ways. For instance one can enforce eq.~\eqref{eq:ren-transf6-bis} within a two-point function. One can consider the wall average of a local observable $\phi_t(x)$:
\begin{gather}
  \Phi_t(x_4) = \frac{1}{L^3} \sum_{\mathbf{x}} a^3 \phi_t(\mathbf{x},x_4)
  \ ,
\end{gather}
and choose the integration volume in eq.~\eqref{eq:ren-transf6-bis} to
be the space-time slice $-d<x_4<d$:
\begin{gather}
  Z_\td \langle  \Phi_t(z_4) \sum_{y_4=-d}^d \sum_{\mathbf{y}} a^4 \hat{\td}_{y,4} \Phi_t(0) \rangle =
  \langle  \Phi_t(z_4) \hat{\partial}_4 \Phi_t(0) \rangle
  + O \left( e^{-\frac{\bar{r}^2}{16t}} \right)
  \ .
  \label{eq:lat-proposal-2}
\end{gather}
The operator $\Phi_t(z_4)$ must lay outside of the integration
slice. The distance $\bar{r}$ that controls the exponential is the
minimum between $d$ and $|z_4-d|$. In order to suppress the
exponential correction, $\bar{r}$ has to be larger than the smearing
range $\sqrt{8t}$. If the exponential is negligible, then there is a
range of values of $d$ for which the l.h.s. of
eq.~\eqref{eq:lat-proposal-2} is constant, and this can be easily
checked in a numerical calculation.

\bigskip

An alternative method, based on the same idea, consists in using a
Schr\"odinger functional setup, with boundaries at $x_4=\pm L_4$. One
needs to engineer boundary conditions such that the background field
$F_{\mu\nu}$ depends on the coordinate $x_4$. In this case
eq.~\eqref{eq:ren-transf6-bis} becomes:
\begin{gather}
  Z_\td \langle  \sum_{y_4=-d}^d \sum_{\mathbf{y}} a^4 \hat{\td}_{y,4} \Phi_t(0) \rangle_{SF} =
  \langle  \hat{\partial}_4 \Phi_t(0) \rangle_{SF}
  + O \left( e^{-\frac{\bar{r}^2}{16t}} \right)
  \ .
  \label{eq:lat-proposal-3}
\end{gather}
The distance $\bar{r}$ that controls the exponential is the minimum between $d$ and $L_4-d$. The advantage of this approach is that only 1-point functions need to be considered.

\bigskip

Finally one can decide to fix the multiplicative renormalization by
means of the DWI~\eqref{eq:CS}. However one has to take into account
corrections coming from the compact geometry of the space-time in
lattice simulations:
\begin{gather}
  \left( 2 t \frac{d}{dt} + d_\phi\right) \langle \phi_t \rangle =
  \left( c_2 + \frac{c_3}{4} \right) \langle \phi_t(0) \sum_{x_i=-d}^d
  a^4 T^{(2)}_{\mu\mu}(x) \rangle_c + O \left(
    e^{-\frac{\bar{r}^2}{16t}} \right) \ .
\end{gather}
The distance $\bar{r}$ that controls the exponential is the minimum
between $d$ and $L-d$.

\section{Remarks on small flow-time expansion}
\label{sec:small-time}

As already shown in ref.~\cite{Suzuki:2013gza}, one can obtain the
energy momentum tensor from the small flow-time expansion of the
following two operators:
\begin{gather}
  Y_{\mu\rho}(t,x) = - 2 \left\{ \tr G_{\sigma\mu}(t,x)
    G_{\sigma\rho}(t,x) - \frac{\delta_{\mu\rho}}{4} \tr
    G_{\sigma\tau}(t,x) G_{\sigma \tau}(t,x) \right\} \ ,
  \\
  E(t,x) = - \frac{1}{2} \tr G_{\sigma\tau}(t,x) G_{\sigma \tau}(t,x)
  \ .
  \label{eq:suzuki-def}
\end{gather}
The small flow-time expansions of the operators $E(t,x)$ and
$Y_{\mu\rho}(t,x)$ are organized in terms of the dimension $d_{k}$ of
the possible mixing renormalized boundary operators $\Theta^{(k)}_R$:
\begin{gather}
  Y_{\mu\rho}(t,x) = \alpha_Y(t) \left[ T_{\mu \rho}(x) -
    \frac{\delta_{\mu\rho}}{4} T_{\sigma\sigma}(x) \right] +
  \sum_{d_{Y,k} \ge 6} t^{d_{Y,k}/2-2} c_{Y,k}(t;\mu)
  \Theta^{(Y,k)}_{R,\mu\nu}(x;\mu) \ ,
  \label{eq:ope_Y}
  \\
  E(t,x) = \langle E(t,x) \rangle + \alpha_E(t) T_{\mu \mu}(x) +
  \sum_{d_{E,k} \ge 6} t^{d_{E,k}/2-2} c_{E,k}(t;\mu)
  \Theta^{(E,k)}_R(x;\mu) \ .
  \label{eq:ope_E}
\end{gather}

The coefficients $\alpha_Y(t)$ and $\alpha_E(t)$ are renormalization group invariant, and have a pertubative expansion in terms of the running coupling $g(q)$ at the energy scale $q=(8t)^{-1/2}$.
As calculated in ref.~\cite{Suzuki:2013gza}:
\begin{gather}
\alpha_Y(t) = g_{MS}^2(q) + 2 b_0 \left[ \ln \sqrt{\pi} + \frac{7}{16} \right] g_{MS}^4(q) + O(g_{MS}^6(q)) \ ,
\\
\alpha_E(t) = \frac{1}{2 b_0} + \left( \frac{109}{176} - \frac{b_1}{2b_0^2} \right) g_{MS}^2(q) + O(g_{MS}^4(q)) \ .
\end{gather}
The calculation is done in the MS renormalization scheme. $b_0$ and $b_1$ are the coefficients of the expansion of the beta function:
\begin{gather}
\beta(g) = -b_0 g^3 - b_1 g^5 + O(g^7) \ .
\end{gather}
The small flow-time behaviour of the coefficients $c_k(t;\mu)$ is dictated by the renormalization group equation. These coefficients are at most logarithmically divergent.\footnote{
The coefficients $c_k$ are dimensionless and depend on the running coupling constant $g(\mu)$ and on the ratio $q/\mu$. They satisfy a renormalization-group equation:
\begin{gather}
\left[ \delta^{jk} q \frac{\partial}{\partial q} - \delta^{jk} \beta(g) \frac{\partial}{\partial g} - \gamma^{jk}(g) \right] c_j = 0 \ ,
\\
\gamma^{jk}(g) = \gamma^{jk}_0 g^2 + O(g^4) \ ,
\end{gather}
where the anomalous dimension matrix accounts for mixing of operators under renormalization group. The leading behaviour of $c_k(t;\mu)$ at large $q$ (i.e. small $t$) is governed by the leading term of the anomalous dimension at small $g$:
\begin{gather}
c_k(t;\mu) \simeq [ g(q)^{-\gamma_0/b_0} ]_{kj} v_j \ ,
\end{gather}
for some vector $v$ that depends on the initial condition of the renormalization group equation. This shows that $c_k(t;\mu)$ can have at most a logarithmic divergence in $t$.

}

\bigskip

We propose here a strategy to determine the coefficients $\alpha_Y(t)$
and $\alpha_E(t)$ nonperturbatively up to $O(t)$ corrections. Let us
focus on the trace of the energy momentum for now. Given a probe
observable $\phi_T$ at positive flow time $T$, one can define an
effective coefficient $\alpha^\mathrm{eff}_E(t)$ by imposing that only the
leading term of OPE contributes to $E(t,x)$:
\begin{gather}
  \langle \phi_T \int d^Dx \ E(t,x) \rangle_c = \alpha^\mathrm{eff}_E(t)
  \langle \phi_T \int d^Dx \ T_{\mu \mu}(x) \rangle \ .
\end{gather}
Notice that $\alpha^\mathrm{eff}_E(t)$ defined in this way depends on the probe observable. By using eq.~\eqref{eq:ope_E}, one easily sees that:
\begin{gather}
\alpha^\mathrm{eff}_E(t)  = \alpha_E(t) + O(t) \ ,
\end{gather}
therefore the nonuniversal terms are at least $O(t)$. Here $O(t)$ has to be interpreted up to logarithmic corrections. By using the dilatation Ward identity in the form of eq.~\eqref{eq:CS}, one obtains the explicit representation:
\begin{gather}
  \label{eq:alphaeffcomput}
  \alpha^\mathrm{eff}_E(t) \left( 2 T \frac{d}{dT} + d_\phi\right)
  \langle \phi_T \rangle =
  \langle \phi_T \int d^Dx \ E(t,x) \rangle_c
  \ .
\end{gather}
In a region in which the nonuniversal $O(t)$ contribution are small, then the following operatorial relation holds:
\begin{gather}
T_{\mu \mu}(x) = \frac{1}{\alpha^\mathrm{eff}_E(t)} \left[ E(t,x) - \langle E(t,x) \rangle \right] + O(t) \ .
\end{gather}
It is worth to stress that this nonperturbative definition of
$\alpha^\mathrm{eff}_E(t)$ provides a definition of the trace of the
energy-momentum tensor that is correct up to order $t$ (times
logarithms), while a perturbative definition of
$\alpha^\mathrm{eff}_E(t)$ as it is pursued in
ref.~\cite{Suzuki:2013gza} would give rise or errors that are of order
$\ln^{-2n} t$.

\bigskip

Analogously one can define an effective coefficient $\alpha^\mathrm{eff}_Y(t)$ by imposing the integrated TWI, for instance in the form:
\begin{gather}
\langle \phi_T(x) \phi_T(0) \int d^{D-1}\mathbf{y} \ [ Y_{Dk}(t,\mathbf{y},d) - Y_{Dk}(t,\mathbf{y},-d) ] \rangle
+ O \left( e^{- \frac{\bar{r}^2}{16T}} \right)
= \nonumber \\
\qquad =
\alpha^\mathrm{eff}_Y(t) \langle \phi_T(x) \partial_k \phi_T(0) \rangle
 \ .
\end{gather}
In this formula we have separated space and time coordinates $x = (\mathbf{x}, x_D)$. The local TWI~\eqref{eq:TWI3} has been integrated in a space-time slice $-d < x_D < d$. The index $k$ runs from $1$ to $D-1$, the point $x$ lays outside of the integration volume, and $\bar{r}$ is the minimum between $d$, $|x_D + d|$ and $|x_D - d|$. Also in this case, by using eq.~\eqref{eq:ope_Y}, one sees that:
\begin{gather}
\alpha^\mathrm{eff}_Y(t)  = \alpha_Y(t) + O(t) \ ,
\\
T_{\mu \rho}(x) - \frac{\delta_{\mu\rho}}{4}T_{\sigma\sigma}(x) = \frac{1}{\alpha^\mathrm{eff}_Y(t)} Y_{\mu\rho}(t,x) + O(t) \ .
\end{gather}

\section{Conclusions}
\label{sec:concl}

Since the renormalization properties of the gradient flow have been
clarified, the latter provides a theoretically robust way to
investigate the dynamics of gauge theories, and several interesting
applications have already appeared since it was originally
introduced. In this paper we focus on the possibility of using the
gradient flow for studying space-time symmetries like translations and
dilatations. An important corollary of our study is that the gradient
flow can be used to define a properly-normalized energy-momentum
tensor for pure Yang-Mills theories defined on a lattice.

The main idea used in this work, inspired by the study in
ref.~\cite{Luscher:2013cpa}, is that the variations under
infinitesimal local translations of correlators of fields along the
flow can be used to generate translation Ward identities, which
encode the symmetry properties of the quantum field theory. We have
explored two applications of this idea.

First we studied the case where the transformation is defined on the
fields at flow time $t=0$, and we obtained the Ward identities using
probe operators at positive flow time. The divergencies of the
correlators appearing in these identities have been analysed using a
representation of the gradient flow in terms of a $(D+1)$-dimensional
local field theory. When a lattice regulator is used,
translation symmetry is broken by the regulator, and the
energy-momentum tensor undergoes renormalization. A finite
energy-momentum tensor can only be defined after the subtraction of
divergent mixings with other operators. The Ward identities for the
renormalized lattice energy-momentum tensor using probe operators at
time $T>0$ are shown in eq.~\eqref{eq:lat-TWI-t}; the key feature is
that these identities can be used to fix the renormalization
coefficients in a nonperturbative way.  These results extend the
programme that was first laid out in refs.~\cite{Caracciolo:1989pt} to
the case of probe operators smeared using the gradient flow. Numerical
simulations are needed to verify that this is a viable method in
practice; they are deferred to future investigations.

Because the gradient flow commutes with uniform translations, we can
also study the Ward identities obtained by transforming the fields at
nonvanishing flow time $t$.  Once again a $(D+1)$-dimensional
representation of the gradient flow allows us to analyse the structure
of the field correlators in terms of local fields in the bulk. We have
obtained the renormalized Ward identities that are generated by these
transformations. They are universal properties of the field
correlators, reflecting the translation invariance of the physical
world, and do not depend on the regulator used to define the bare
theory. The Noether currents appearing in these Ward identities are
related to the energy-momentum tensor of the original $D$-dimensional
theory in eq.~\eqref{eq:TWI3}.

Our analysis includes the case of dilatations. Indeed local
dilatations are a special case of local translations. Studying local
dilatations in the bulk, we were able to write dilatation Ward
identities for operators at generic flow time $T$. These Ward
identities show explicitly the anomalous breaking of scale invariance,
and thereby provide a new tool to study the trace of the
energy-momentum tensor. The variation of the probe fields along the
gradient flow is directly related to the correlator of the trace of
the energy-momentum tensor with the probe fields, as shown in
eq.~\eqref{eq:CS}. This is a remarkable result that allows the
scale invariance of the theory to be probed using the gradient flow.

An interesting extension of the results in ref.~\cite{Suzuki:2013gza}
emerges naturally in the framework used here to discuss the
transformation properties under dilatations. In
ref.~\cite{Suzuki:2013gza} the $D$-dimensional energy-momentum tensor
was defined using a perturbative determination of the small flow-time
expansion of operators defined in the bulk. It is possible to
introduce a nonperturbative definition of the leading coefficients in
this expansion, and to compute them making use of probe observables
along the gradient flow, see e.g. eq.~\ref{eq:alphaeffcomput}.

Using the gradient flow to study space-time symmetries is a fertile
research direction. The recent extension of the gradient flow to
theories with fermions~\cite{Luscher:2013cpa} should enable a
straightforward generalization of our arguments to gauge theories
coupled to matter. We plan to come back on these topics in future
studies.

\acknowledgments We are indebted to Martin L\"uscher for insightful
comments at all stages of this work. AP would like to thank him for
enlightening discussions and constant inspiration. We would like to
thank Roman Zwicky for critical discussions of the manuscript. Part of this work has been discussed and developed during the ``Strongly interacting dynamics beyond the Standard Model and the Higgs boson'' hosted by the \textit{Higgs Centre for Theoretical Physics} in Edinburgh, UK.

\pagebreak

\appendix

\section{Integration of the Lagrange multiplier}
\label{app:L}

When observables depend linearly on the Lagrange multiplier $L_\mu$, like in the case of the $\tilde{T}_{MR}$ defined in eqs.~\eqref{eq:T_0R} and \eqref{eq:T_nR}, the field $L_\mu$ can be explicitly integrated out. We consider functional integrals of the particular form:
\begin{gather}
\int \mathcal{D}B \mathcal{D}L \ P \ X^A(t,x) L^A_\mu(t,x) e^{-S_\mathrm{bulk}} \ ,
\label{eq:app-1}
\end{gather}
where both $P$ and $X^A$ are functions of the field $B$ only, and moreover $X^A$ is a local function of the field $B$ and its spatial derivatives in the point $(t,x)$ only.

\bigskip

For the calculation it is convenient to consider the flow-time direction discretized with a time step equal to $\Delta t$ which we will send to zero at the end of the calculation. We remind that in the notations of~\cite{Luscher:2011bx} the field $L^A_\mu$ is imaginary. The discretized bulk action is:
\begin{gather}
S_\mathrm{bulk} = \sum_{t=0}^T \Delta t \int d^Dx \ L^A_\mu(t,x) \mathcal{F}^A_\mu[B](t,x) \ ,
\\
\mathcal{F}_\mu[B](t,x) = \frac{1}{\Delta t} \left[ B_\mu(t+\Delta t,x) - B_\mu(t,x) \right] - D_\nu G_{\nu\mu}(t,x) \ .
\end{gather}
The integration measure $\mathcal{D}L$ is normalized in such a way that:
\begin{gather}
\int \mathcal{D}L \ e^{-S_\mathrm{bulk}} = \prod_{t=0}^T \prod_x \delta(\mathcal{F}[B](t,x)) = \mathcal{N} \prod_{t=\Delta t}^{T+\Delta t} \prod_x  \delta(B(t,x)-\bar{B}_t(x)) \ ,
\end{gather}
where the field $\bar{B}_t(x)$ is the solution of the discretized gradient flow equation with initial condition $\bar{B}_0 = A$. The second equality in the previous equation is obtained by changing variables from $[\mathcal{F}(t)]_{t=0,\dots,T}$ to $[B(t)]_{t=\Delta t,\dots,T+\Delta t}$ (the shift in the indices is important!). The Jacobian matrix of this map is (proportional to):
\begin{flalign}
\Delta^{AB}_{\mu\nu}[B](t,x;s,y)
& \overset{\mathrm{def}}{=}
\frac{1}{\Delta t}
\frac{\delta \mathcal{F}^A_\mu(t,x)}{\delta B^B_\nu(s,y)} = \nonumber\\
& \overset{\phantom{\mathrm{def}}}{=}
\frac{\delta_{t+\Delta T, s} - \delta_{t, s}}{\Delta t^2} \delta^{AB} \delta_{\mu\nu} \delta(x-y) - \frac{\delta_{t, s}}{\Delta t} R^{AB}_{\mu\nu}(t;x,y)
\ ,
\label{eq:app-2}
\end{flalign}
where the matrix $R$ is defined as:
\begin{gather}
\frac{\delta D_\rho G_{\rho\mu}^A(t,x)}{\delta B^B_\nu(s,y)} = R^{AB}_{\mu\nu}(t;x,y) \delta_{t,s} \ .
\end{gather}
Notice that $\Delta[B]$ is an upper triangular matrix, whose diagonal is in $s=t+\Delta T$. Its determinant (which requires regularization) is just the product of all the diagonal entries of the matrix $\Delta[B]$ and does not depend on the fields~\cite{ZinnJustin:1986eq, Luscher:2011bx}. 

\bigskip

Let us isolate the functional integral in $L_\mu$ from eq.~\eqref{eq:app-1}:
\begin{gather}
\int \mathcal{D}L \ L^A_\mu(t,x) e^{-S_\mathrm{bulk}}
=
- \frac{1}{\Delta t} \frac{\delta}{\delta \mathcal{F}^A_\mu(t,x)} \int \mathcal{D}L \ e^{-S_\mathrm{bulk}}
= 
\nonumber
\\ =
- \frac{1}{\Delta t} \frac{\delta}{\delta \mathcal{F}^A_\mu(t,x)} \delta(\mathcal{F}[B])
=
- \frac{\mathcal{N}}{\Delta t} \sum_{s=\Delta t}^{T+\Delta t} \Delta t \int d^Dy \  \frac{\delta B^B_\nu(s,y)}{\delta \mathcal{F}^A_\mu(t,x)} \frac{\delta}{\delta B^B_\nu(s,y)} \delta(B-\bar{B}) \ .
\label{eq:app-3}
\end{gather}
When this result is plugged back into the original integral~\eqref{eq:app-1}, one can integrate by part in $B$ and get
\begin{flalign}
\int \mathcal{D}B & \mathcal{D}L \ P \ X^A(t,x) L^A_\mu(t,x) \ e^{-S_\mathrm{bulk}} 
=
\label{eq:monster-1}
\\
= & \frac{\mathcal{N}}{\Delta t} \sum_{s=\Delta t}^{T+\Delta t} \Delta t \int d^Dy \ 
\left[\frac{\delta P}{\delta B^B_\nu(s,y)}
\frac{\delta B^B_\nu(s,y)}{\delta \mathcal{F}^A_\mu(t,x)}
X^A(t,x)
\right]_{B=\bar{B}}
+
\label{eq:monster-2}
\\
& +
\mathcal{N} \int d^Dy \ 
\left[
P
\frac{\delta B^B_\nu(t,y)}{\delta \mathcal{F}^A_\mu(t,x)}
\frac{\delta X^A(t,x)}{\delta B^B_\nu(t,y)}
\right]_{B=\bar{B}}
+
\label{eq:monster-3}
\\
& +
\frac{\mathcal{N}}{\Delta t} \sum_{s=\Delta t}^{T+\Delta t} \Delta t \int d^Dy \ 
\left[
P
X^A(t,x)
\frac{\delta}{\delta B^B_\nu(s,y)} \frac{\delta B^B_\nu(s,y)}{\delta \mathcal{F}^A_\mu(t,x)}
\right]_{B=\bar{B}} \ .
\label{eq:monster-4}
\end{flalign}
We will show that the terms~\eqref{eq:monster-3} and \eqref{eq:monster-4} vanish.

\bigskip

The matrix $\delta B / \delta \mathcal{F}$ appearing in the previous equation is the inverse of the $\Delta[B]$ matrix defined in eq.~\eqref{eq:app-2}. Moreover as $\Delta[B]$ is an upper triangular matrix, also $\delta B / \delta \mathcal{F}$ is an upper triangular matrix. The equations that define $\delta B / \delta \mathcal{F}$ are:
\begin{flalign}
\hspace{5mm} & \frac{1}{\Delta t} \frac{\delta B^C_\rho(s,y)}{\delta \mathcal{F}^A_\mu(t,x)} = 0
\ ,
&& s \le t
\ ,
\label{eq:app-4a}
\\
& \frac{1}{\Delta t} \frac{\delta B^C_\rho(s,y)}{\delta \mathcal{F}^A_\mu(t,x)} = \delta^{CA} \delta_{\rho\mu} \delta(y-x)
\ ,
&& s = t+\Delta t
\ ,
\label{eq:app-4b}
\\
& \int d^Dz \ \left[ \delta^{BC} \delta_{\nu\rho} \delta(y-z) \hat{\partial}_s^+ - R^{BC}_{\nu\rho}(s;y,z) \right] \frac{1}{\Delta t} \frac{\delta B^C_\rho(s,z)}{\delta \mathcal{F}^A_\mu(t,x)}
=
0
\ ,
&& s > t+\Delta t
\ ,
\label{eq:app-4c}
\end{flalign}
where $\hat{\partial}_s^+$ is the discrete forward derivative that appears in eq.~\eqref{eq:app-2}. Notice that eq.~\eqref{eq:app-4a} implies that the term~\eqref{eq:monster-3} vanishes.

The solution of the eqs.~\eqref{eq:app-4a}, \eqref{eq:app-4b}, \eqref{eq:app-4c} can be found iteratively (for $s > t+\Delta t$):
\begin{gather}
\frac{1}{\Delta t} \frac{\delta B^B_\nu(s,y)}{\delta \mathcal{F}^A_\mu(t,x)}
=
\int d^Dz \ \left[ \delta^{BC} \delta_{\nu\rho} \delta(y-z) + \Delta t R^{BC}_{\nu\rho}(s-1;y,z) \right] \frac{1}{\Delta t} \frac{\delta B^C_\rho(s-1,z)}{\delta \mathcal{F}^A_\mu(t,x)} \ .
\label{eq:app-5}
\end{gather}
From here it is clear that $\delta B^B_\nu(s,y) / \delta \mathcal{F}^A_\mu(t,x)$ does not depend on the field $B$ at flow time $s$, therefore the term~\eqref{eq:monster-4} vanishes.

Finally we want to notice that:
\begin{gather}
\left[
\frac{1}{\Delta t} \frac{\delta B^B_\nu(s,y)}{\delta \mathcal{F}^A_\mu(t,x)}
\right]_{B=\bar{B}}
=
\frac{\delta \bar{B}^B_{s,\nu}(y)}{\delta \bar{B}^A_{t+\Delta t,\mu}(x)} \ ,
\label{eq:app-6}
\end{gather}
for $s \ge t+\Delta t$, where the r.h.s. is the Jacobian matrix of the map $\bar{B}_{t+\Delta t} \mapsto \bar{B}_s$. In fact the Jacobian matrix $\delta \bar{B}^B_{s,\nu}(y) / \delta \bar{B}^A_{t+\Delta t,\mu}(x)$ satisfies a (discretized) differential equation that is obtained by differentiating the discretized flow equation. A rapid inspection shows that this differential equation coincides with eq.~\eqref{eq:app-4c} after the substitution $B=\bar{B}$. Also both matrices in eq.~\eqref{eq:app-6} satisfy the same initial condition~\eqref{eq:app-4b}. eq.~\eqref{eq:app-6} follows from the uniqueness of the solution of eq.~\eqref{eq:app-4c}.

\bigskip

By taking the $\Delta t \to 0$ limit, eq.~\eqref{eq:monster-1} becomes then:
\begin{flalign}
&\int \mathcal{D}B \mathcal{D}L \ P \ X^A(t,x) L^A_\mu(t,x) \ e^{-S_\mathrm{bulk}} 
= \nonumber \\
& \qquad =
\mathcal{N} \int_{t}^{\infty} ds \int d^Dy \ 
\left[\frac{\delta P}{\delta B^B_\nu(s,y)}
\frac{\delta \bar{B}^B_{s,\nu}(y)}{\delta \bar{B}^A_{t,\mu}(x)}
X^A(t,x)
\right]_{B=\bar{B}} \ ,
\label{eq:app-main-1}
\end{flalign}
which is the main formula of this appendix. Notice that in the $\Delta t \to 0$ limit one has to replace:
\begin{gather}
\frac{1}{\Delta t} \frac{\delta}{\delta B^B_\nu(s,y)} \to \frac{\delta}{\delta B^B_\nu(s,y)} \ ,
\end{gather}
as $s$ becomes a continuous parameter.

If $P_T$ is function of the field $B$ at some positive flow time $T>t$ only, then the functional derivative with respect to $B(s,y)$ contains a delta function that can be extracted:
\begin{gather}
\frac{\delta P_T}{\delta B^B_\nu(s,y)} = \frac{\delta P_T}{\delta \bar{B}^B_{T,\nu}(y)} \delta(s-T) \ .
\end{gather}
Plugging this into eq.~\eqref{eq:app-main-1} yields:
\begin{flalign}
&\int \mathcal{D}B \mathcal{D}L \ P_T \ X^A(t,x) L^A_\mu(t,x) \ e^{-S_\mathrm{bulk}} 
= \nonumber \\
& \qquad =
\mathcal{N} \int d^Dy \ 
\left[\frac{\delta P_T}{\delta \bar{B}^B_{T,\nu}(y)}
\frac{\delta \bar{B}^B_{s,\nu}(y)}{\delta \bar{B}^A_{t,\mu}(x)}
X^A(t,x)
\right]_{B=\bar{B}}
= \nonumber \\
& \qquad =
\mathcal{N}
\frac{\delta P_T[\bar{B}_T]}{\delta \bar{B}^A_{t,\mu}(x)}
\left[ X^A(t,x) \right]_{B=\bar{B}} \ ,
\label{eq:app-main-2}
\end{flalign}
where the chain rule has been used in the last step. This equation has been used several times in this paper. Equation~\eqref{eq:variation-1} is a particular instance of the previous equation.

\bigskip

Another interesting class of functional integrals is represented by 
\begin{gather}
\int \mathcal{D}B \mathcal{D}L \ P \  \mathcal{F}^A_\nu(t,x) \ L^A_\mu(t,x) e^{-S_\mathrm{bulk}} = 0\ ,
\label{eq:app-f1}
\end{gather}
where again $P$ is function of the field $B$ only. We have used many
times in this paper the fact that this integral vanishes. Moreover
this integral is useful to extend, by subtraction, the result in
eq.~\eqref{eq:app-main-2} to observables $X^A$ that include a
flow-time derivative of the field $B$. We can follow the previous
derivation and use \eqref{eq:app-1} to obtain
\begin{flalign}
\int \mathcal{D}B & \mathcal{D}L \ P \ \mathcal{F}^A_\nu(t,x) L^A_\mu(t,x) \ e^{-S_\mathrm{bulk}} 
=
\label{eq:monster-f1}
\\
= & \frac{\mathcal{N}}{\Delta t} \sum_{s=\Delta t}^{T+\Delta t} \Delta t \int d^Dy \ 
\left[\frac{\delta P}{\delta B^B_\nu(s,y)}
\frac{\delta B^B_\nu(s,y)}{\delta \mathcal{F}^A_\mu(t+\Delta t,x)}
\mathcal{F}^A_\nu(t,x) 
\right]_{B=\bar{B}}
+
\label{eq:monster-f2}
\\
& +
\left[
P
\frac{\delta \mathcal{F}^A_\nu (t+\Delta t,x)}{\delta \mathcal{F}^A_\mu(t+\Delta t,x)}
\right]_{B=\bar{B}}
+
\label{eq:monster-f3}
\\
& +
\frac{\mathcal{N}}{\Delta t} \sum_{s=\Delta t}^{T+\Delta t} \Delta t \int d^Dy \ 
\left[
P
\mathcal{F}^A_\nu(t,x)
\frac{\delta}{\delta B^B_\nu(s,y)} \frac{\delta B^B_\nu(s,y)}{\delta \mathcal{F}^A_\mu(t+\Delta t,x)}
\right]_{B=\bar{B}} \ .
\label{eq:monster-f4}
\end{flalign}

Since  the terms~\eqref{eq:monster-f2} and \eqref{eq:monster-f4} are
linear in $\mathcal{F}$, they vanish once evaluated on the
constraint. The term ~\eqref{eq:monster-f3} is proportional to a
$\delta(0)$ that in dimensional regularization is automatically
regularized to zero.

\section{Integration of the Lagrange multiplier on the lattice}
\label{app:L-lat}

Equation~\eqref{eq:app-main-1} can be generalized to the lattice. We sketch here the calculation. The four space-time coordinates are discretized with a lattice spacing $a$, and the flow-time coordinate is discretized with a lattice spacing $\Delta t$. The bulk action is (see ref.~\cite{Luscher:2013cpa} for details):
\begin{flalign}
S_{\mathrm{bulk}} = & \sum_{t=\Delta t}^{T} \Delta t \sum_{x, \mu, A} a^4
L^A_\mu(t-\Delta t,x) \frac{1}{a \Delta t} \hat{\mathcal{F}}^A_\mu[V](t,x)
+ \nonumber \\
& \qquad
- \sum_{t=\Delta t}^{T} \sum_{x,\mu} \ln \det K(V_\mu(t,x) V^\dag_\mu(t-\Delta t,x))
\ .
\label{eq:app-2-bulk}
\end{flalign}
The Lagrange multiplier $L$ and the functional $\hat{\mathcal{F}}_\mu$ live in the Lie algebra of the gauge group $SU(N)$, and the equation $\hat{\mathcal{F}}_\mu = 0$ generates the flow equation. Explicitly:
\begin{gather}
\hat{\mathcal{F}}_\mu[V](t,x) = \mathcal{P} \left[ V_\mu(t,x) V^\dag_\mu(t-\Delta t,x) - e^{- \Delta t g_0^2 \partial S_W[V](t-\Delta t,x,\mu)} \right] \ ,
\label{eq:app-2-constraint}
\end{gather}
where $\mathcal{P}$ is the projector on the Lie algebra:
\begin{gather}
\mathcal{P}(U) = \frac{U-U^\dag}{2} -\frac{1}{N} \tr \frac{U-U^\dag}{2} \ .
\end{gather}
When restricted to a neighborhood of the identity in the gauge group, the projector $\mathcal{P}$ is invertible and its Jacobian is:
\begin{gather}
K_{AB}(U) = -2 \partial^B \tr T^A \mathcal{P}(U) = -2 \tr T^A \mathcal{P} (T^B U) \ .
\end{gather}
Throughout this paper, the Lie derivative on the gauge group is defined as:
\begin{gather}
\partial^A f(U) = \left. \frac{d}{d\alpha} \right|_{\alpha =0} f(e^{\alpha T^A} U) \ .
\end{gather}
The functional integral over the bulk field $V$ is restricted on a neighborhood of the solution of the flow equation in which the map $V \mapsto \hat{\mathcal{F}}$ is invertible. As we do not need to know this domain explicitly, we will omit it in the next formulae. By inspecting eq.~\eqref{eq:app-2-constraint} one sees that $\hat{\mathcal{F}}_\mu(t,x)$ depends on the field $V_\nu(s,y)$ at flow times $s \le t$ only. The Jacobian matrix of the map $V \mapsto \hat{\mathcal{F}}$ is block-triangular, and its determinant is given by the product of the determinants of the diagonal blocks:
\begin{gather}
\prod_{t=\Delta t}^{T} \prod_{x,\mu} \det_{AB} [ \partial^A_{V_\mu(t,x)} \hat{\mathcal{F}}^B_\mu(t,x) ]
=
\prod_{t=\Delta t}^{T} \prod_{x,\mu} \det K(V_\mu(t,x) V^\dag_\mu(t-\Delta t,x)) \ .
\end{gather}
This Jacobian determinant is at the origin of the extra piece in the bulk action in eq.~\eqref{eq:app-2-bulk}.

\bigskip

Integrals of the form:
\begin{gather}
\mathcal{I} = \int \mathcal{D}V \mathcal{D}L \ P_T[V] \ X^A[V](t,x) L^A_\mu(t,x) \ e^{-S_\mathrm{bulk}}
\end{gather}
can be easily calculated by using the change of variable $V \mapsto \hat{\mathcal{F}}$:
\begin{flalign}
\mathcal{I} 
= & - \int \mathcal{D}\hat{\mathcal{F}} \mathcal{D}L \ P_T[V[\hat{\mathcal{F}}]] \ X^A[V[\hat{\mathcal{F}}]](t,x) \times \nonumber \\
& \qquad \qquad \times
\frac{\partial}{\partial \hat{\mathcal{F}}^A_\mu(t+\Delta t,x)} \ e^{ - \sum_{s, y, \nu, B} a^3
L^B_\nu(s-\Delta t,y) \hat{\mathcal{F}}^B_\nu(s,y)
}
= \nonumber \\
= &
\left[ \frac{\partial}{\partial \hat{\mathcal{F}}^A_\mu(t+\Delta t,x)}  P_T[V[\hat{\mathcal{F}}]] \ X^A[V[\hat{\mathcal{F}}]](t,x) \right]_{\hat{\mathcal{F}}=0} \ .
\end{flalign}
We notice now that, if $X^A(t,x)$ depends on bulk fields at flow time $t$ only, then it does not depend on $\hat{\mathcal{F}}$ at flow time $t+\Delta t$. Also we will assume that $P_T$ depends on the fields at flow time $T$ only. The constraint $\mathcal{F}=0$ is equivalent to requiring that the bulk field satisfies the flow equation $V=\bar{V}$:
\begin{gather}
\mathcal{I} 
=
-2 X^A[\bar{V}](t,x) \sum_{y,\nu,B} \tr \left[ T^B \frac{\partial V_\nu(T,y)}{\partial \hat{\mathcal{F}}^A_\mu(t+\Delta t,x)} V^\dag_\nu(T,y) \right]_{\hat{\mathcal{F}}=0} \partial^B_{V_\nu(T,y)} P_T[\bar{V}] \   \ .
\label{eq:app-2-I5}
\end{gather}
We want to argue now that the partial derivative in the previous equation is related to the Jacobian matrix of the trivializing map $\bar{V}_{t+\Delta t} \mapsto \bar{V}_T$. In order to do so it is convenient to introduce the following differential:
\begin{gather}
dB^B_\nu(s,y) = -2 \tr [ V_\nu(s,y) T^B \, dV_\nu(s,y) ] \ ,
\end{gather}
in terms of which the Jacobian matrix of the trivializing map in the $\Delta t \to 0$ limit is:
\begin{flalign}
J^{BA}_{\nu\mu}(T,y;t,x) = & \lim_{\Delta t \to 0} \frac{1}{a^4} \left[ \frac{\partial B^B_\nu(T,y)}{\partial B^A_\mu(t+\Delta t,x)} \right]_{\hat{\mathcal{F}}=0}
= \nonumber \\
= &
\lim_{\Delta t \to 0} \frac{1}{a^4}
\sum_{z,\rho,C}
\left[
\frac{\partial B^B_\nu(T,y)}{\partial \hat{\mathcal{F}}^C_\rho(t+\Delta t,z)} 
\frac{\partial \hat{\mathcal{F}}^C_\rho(t+\Delta t,z)}{\partial B^A_\mu(t+\Delta t,x)}
\right]_{\hat{\mathcal{F}}=0}
= \nonumber \\
= &
\lim_{\Delta t \to 0} \frac{1}{a^4}
\left[ \frac{\partial B^B_\nu(T,y)}{\partial \hat{\mathcal{F}}^C_\mu(t+\Delta t,x)} \right]_{\hat{\mathcal{F}}=0}
K_{CA}(\bar{V}_\mu(t+\Delta t,x) \bar{V}^\dag_\mu(t,x))
= \nonumber \\
= &
\lim_{\Delta t \to 0} \frac{1}{a^4}
\left[ \frac{\partial B^B_\nu(T,y)}{\partial \hat{\mathcal{F}}^A_\mu(t+\Delta t,x)} \right]_{\hat{\mathcal{F}}=0}
\ ,
\end{flalign}
as $K(\mathbf{1})$ is the identity matrix. Plugging this formula back in eq.~\eqref{eq:app-2-I5} in the $\Delta t \to 0$ we get the desired result:
\begin{flalign}
& \int \mathcal{D}V \mathcal{D}L \ P_T[V] \ X^A[V](t,x) L^A_\mu(t,x) \ e^{-S_\mathrm{bulk}} =
\nonumber \\
& \qquad =
X^A[\bar{V}](t,x) \sum_{y,\nu,B} a^4 J^{BA}_{\nu\mu}(T,y;t,x) \partial^B_{V_\nu(T,y)} P_T[\bar{V}] \ .
\end{flalign}

For instance, if $t=0$ then one can use the chain rule to show that:
\begin{gather}
\int \mathcal{D}V \mathcal{D}L \ P_T[V] \ X^A[V](0,x) L^A_\mu(0,x) \ e^{-S_\mathrm{bulk}} =
X^A[U](x) \, \partial^B_{U_\nu(y)} P_T[\bar{V}] \ ,
\end{gather}
where $U$ is the boundary field. eq.~\eqref{eq:variation-1-lat} is just a particular application of this formula.

\pagebreak

\bibliographystyle{JHEP}
\bibliography{paper}

\providecommand{\href}[2]{#2}\begingroup\raggedright\begin{thebibliography}{10}

\bibitem{Caracciolo:1989pt}
S.~Caracciolo, G.~Curci, P.~Menotti, and A.~Pelissetto, {\it The energy
  momentum tensor for lattice gauge theories},  {\em Annals Phys.} {\bf 197}
  (1990) 119.

\bibitem{Caracciolo:1988hc}
S.~Caracciolo, G.~Curci, P.~Menotti, and A.~Pelissetto, {\it The energy
  momentum tensor on the lattice: The scalar case},  {\em Nucl.Phys.} {\bf
  B309} (1988) 612.

\bibitem{Suzuki:2013gza}
H.~Suzuki, {\it {Energy-momentum tensor from the Yang--Mills gradient flow}},
  \href{http://xxx.lanl.gov/abs/1304.0533}{{\tt arXiv:1304.0533}}.

\bibitem{Suzuki:2013gi}
H.~Suzuki, {\it {Ferrara--Zumino supermultiplet and the energy-momentum tensor
  in the lattice formulation of 4D $\mathcal{N}=1$ SYM}},  {\em Nucl.Phys.}
  {\bf B868} (2013) 459--475, [\href{http://xxx.lanl.gov/abs/1209.2473}{{\tt
  arXiv:1209.2473}}].

\bibitem{Suzuki:2012wx}
H.~Suzuki, {\it {Remark on the energy-momentum tensor in the lattice
  formulation of 4D $\mathcal{N}=1$ SYM}},  {\em Phys.Lett.} {\bf B719} (2013)
  435--439, [\href{http://xxx.lanl.gov/abs/1209.5155}{{\tt arXiv:1209.5155}}].

\bibitem{Luscher:2009eq}
M.~Luscher, {\it {Trivializing maps, the Wilson flow and the HMC algorithm}},
  {\em Commun.Math.Phys.} {\bf 293} (2010) 899--919,
  [\href{http://xxx.lanl.gov/abs/0907.5491}{{\tt arXiv:0907.5491}}].

\bibitem{Luscher:2010iy}
M.~Luscher, {\it {Properties and uses of the Wilson flow in lattice QCD}},
  {\em JHEP} {\bf 1008} (2010) 071,
  [\href{http://xxx.lanl.gov/abs/1006.4518}{{\tt arXiv:1006.4518}}].

\bibitem{Luscher:2011bx}
M.~Luscher and P.~Weisz, {\it {Perturbative analysis of the gradient flow in
  non-abelian gauge theories}},  {\em JHEP} {\bf 1102} (2011) 051,
  [\href{http://xxx.lanl.gov/abs/1101.0963}{{\tt arXiv:1101.0963}}].

\bibitem{Luscher:2013cpa}
M.~Luscher, {\it {Chiral symmetry and the Yang--Mills gradient flow}},  {\em
  JHEP} {\bf 1304} (2013) 123, [\href{http://xxx.lanl.gov/abs/1302.5246}{{\tt
  arXiv:1302.5246}}].

\bibitem{Callan:1970yg}
J.~Callan, Curtis~G., {\it {Broken scale invariance in scalar field theory}},
  {\em Phys.Rev.} {\bf D2} (1970) 1541--1547.

\bibitem{Symanzik:1970rt}
K.~Symanzik, {\it {Small distance behavior in field theory and power
  counting}},  {\em Commun.Math.Phys.} {\bf 18} (1970) 227--246.

\bibitem{Symanzik:1971vw}
K.~Symanzik, {\it {Small distance behavior analysis and Wilson expansion}},
  {\em Commun.Math.Phys.} {\bf 23} (1971) 49--86.

\bibitem{Giusti:2010bb}
L.~Giusti and H.~B. Meyer, {\it {Thermal momentum distribution from path
  integrals with shifted boundary conditions}},  {\em Phys.Rev.Lett.} {\bf 106}
  (2011) 131601, [\href{http://xxx.lanl.gov/abs/1011.2727}{{\tt
  arXiv:1011.2727}}].

\bibitem{Giusti:2012yj}
L.~Giusti and H.~B. Meyer, {\it {Implications of Poincare symmetry for thermal
  field theories in finite-volume}},  {\em JHEP} {\bf 1301} (2013) 140,
  [\href{http://xxx.lanl.gov/abs/1211.6669}{{\tt arXiv:1211.6669}}].

\bibitem{Luscher:2010ik}
M.~Luscher and F.~Palombi, {\it {Universality of the topological susceptibility
  in the SU(3) gauge theory}},  {\em JHEP} {\bf 1009} (2010) 110,
  [\href{http://xxx.lanl.gov/abs/1008.0732}{{\tt arXiv:1008.0732}}].

\bibitem{Luscher:2010we}
M.~Luscher, {\it {Topology, the Wilson flow and the HMC algorithm}},  {\em PoS}
  {\bf LATTICE2010} (2010) 015, [\href{http://xxx.lanl.gov/abs/1009.5877}{{\tt
  arXiv:1009.5877}}].

\bibitem{Fritzsch:2013je}
P.~Fritzsch and A.~Ramos, {\it {The gradient flow coupling in the Schr\"odinger
  Functional}},  \href{http://xxx.lanl.gov/abs/1301.4388}{{\tt
  arXiv:1301.4388}}.

\bibitem{Fodor:2012td}
Z.~Fodor, K.~Holland, J.~Kuti, D.~Nogradi, and C.~H. Wong, {\it {The Yang-Mills
  gradient flow in finite volume}},  {\em JHEP} {\bf 1211} (2012) 007,
  [\href{http://xxx.lanl.gov/abs/1208.1051}{{\tt arXiv:1208.1051}}].

\bibitem{Borsanyi:2012zr}
S.~Borsanyi, S.~Durr, Z.~Fodor, S.~D. Katz, S.~Krieg, et~al., {\it {Anisotropy
  tuning with the Wilson flow}},  \href{http://xxx.lanl.gov/abs/1205.0781}{{\tt
  arXiv:1205.0781}}.

\bibitem{Jackiw:1978ar}
R.~Jackiw, {\it Gauge covariant conformal transformations},  {\em
  Phys.Rev.Lett.} {\bf 41} (1978) 1635.

\bibitem{Berg:2000ak}
B.~A. Berg, {\it {The Transformations of nonAbelian gauge fields under
  translations}},  \href{http://xxx.lanl.gov/abs/hep-th/0006045}{{\tt
  hep-th/0006045}}.

\bibitem{Callan:1970ze}
J.~Callan, Curtis~G., S.~R. Coleman, and R.~Jackiw, {\it {A New improved energy
  - momentum tensor}},  {\em Annals Phys.} {\bf 59} (1970) 42--73.

\bibitem{Adler:1976zt}
S.~L. Adler, J.~C. Collins, and A.~Duncan, {\it {Energy-Momentum-Tensor Trace
  Anomaly in Spin 1/2 Quantum Electrodynamics}},  {\em Phys.Rev.} {\bf D15}
  (1977) 1712.

\bibitem{Coleman:1970je}
S.~R. Coleman and R.~Jackiw, {\it {Why dilatation generators do not generate
  dilatations?}},  {\em Annals Phys.} {\bf 67} (1971) 552--598.

\bibitem{Collins:1976yq}
J.~C. Collins, A.~Duncan, and S.~D. Joglekar, {\it {Trace and Dilatation
  Anomalies in Gauge Theories}},  {\em Phys.Rev.} {\bf D16} (1977) 438--449.

\bibitem{Fujikawa:1980rc}
K.~Fujikawa, {\it {Energy Momentum Tensor in Quantum Field Theory}},  {\em
  Phys.Rev.} {\bf D23} (1981) 2262.

\bibitem{Bochicchio:1985xa}
M.~Bochicchio, L.~Maiani, G.~Martinelli, G.~C. Rossi, and M.~Testa, {\it
  {Chiral Symmetry on the Lattice with Wilson Fermions}},  {\em Nucl.Phys.}
  {\bf B262} (1985) 331.

\bibitem{ZinnJustin:1986eq}
J.~Zinn-Justin, {\it Renormalization and stochastic quantization},  {\em
  Nucl.Phys.} {\bf B275} (1986) 135.

\end{thebibliography}\endgroup

\end{document}